\documentclass[10pt]{article}
\usepackage{epsfig,amsmath,amsfonts,amssymb, cite}
\newcommand{\url}[1]{{\tt #1}}
\newtheorem{theorem}{Theorem}[section]
\newtheorem{lemma}[theorem]{Lemma}

\newtheorem{assumption}[theorem]{Assumption}

\makeatletter
\@addtoreset{equation}{section}
\makeatother

\makeatletter
\long\def\@makecaption#1#2{{\small
\advance\leftskip1cm
\advance\rightskip1cm
\vskip\abovecaptionskip
\sbox\@tempboxa{#1: #2}%
\ifdim \wd\@tempboxa >\hsize
 #1: #2\par
\else
\global \@minipagefalse
\hb@xt@\hsize{\hfil\box\@tempboxa\hfil}%
\fi
\vskip\belowcaptionskip}}
\makeatother
\def\eq#1\en{\begin{equation}#1\end{equation}}  
\def\eqa#1\ena{\begin{align}#1\end{align}}
\def\eqg#1\eng{\begin{gather}#1\end{gather}}
\newcommand{\lb}[1]{\label{e:#1}}
\newcommand{\rlb}[1]{\eqref{e:#1}} 
\newcommand{\nl}{\notag\\}

\newcommand{\snorm}[1]{\Vert#1\Vert}

\newcommand{\bkt}[1]{\left\langle#1\right\rangle}
\newcommand{\sbkt}[1]{\langle#1\rangle}

\newcommand{\sumtwo}[2]%
{\mathop{\sum_{#1}}_{#2}}
\newcommand{\sumthree}[3]%
{\mathop{\mathop{\sum_{#1}}_{#2}}_{#3}}
\newcommand{\sumfour}[4]%
{\mathop{\mathop{\mathop{\sum_{#1}}_{#2}}_{#3}}_{#4}} 
\newcommand{\prodtwo}[2]%
{\mathop{\prod_{#1}}_{#2}}
\newcommand{\mintwo}[2]%
{\mathop{\min_{#1}}_{#2}}
\newcommand{\maxtwo}[2]%
{\mathop{\max_{#1}}_{#2}}
\newcommand{\maxthree}[3]%
{\mathop{\mathop{\max_{#1}}_{#2}}_{#3}}
\newcommand{\limtwo}[2]%
{\mathop{\lim_{#1}}_{#2}}
\newcommand{\suptwo}[2]%
{\mathop{\sup_{#1}}_{#2}}
\newcommand{\supthree}[3]%
{\mathop{\mathop{\sup_{#1}}_{#2}}_{#3}}
\newcommand{\supfour}[4]%
{\mathop{\mathop{\mathop{\sup_{#1}}_{#2}}_{#3}}_{#4}} 
\newcommand{\inftwo}[2]%
{\mathop{\inf_{#1}}_{#2}}
\newcommand{\infthree}[3]%
{\mathop{\mathop{\inf_{#1}}_{#2}}_{#3}}
\newcommand{\inffour}[4]%
{\mathop{\mathop{\mathop{\inf_{#1}}_{#2}}_{#3}}_{#4}} 

\newcommand\calE{{\cal E}}

\newcommand\calH{{\cal H}}

\newcommand\calK{{\cal K}}

\newcommand\calN{{\cal N}}

\newcommand\calS{{\cal S}}

\newcommand{\bsk}{\boldsymbol{k}}

\newcommand{\bsn}{\boldsymbol{n}}

\newcommand{\bsx}{\boldsymbol{x}}

\newcommand{\bspsi}{\boldsymbol{\psi}}

\newcommand{\bbC}{\mathbb{C}}

\newcommand{\bbQ}{\mathbb{Q}}
\newcommand{\bbR}{\mathbb{R}}
\newcommand{\bbZ}{\mathbb{Z}}
\newcommand{\ep}{\varepsilon}
\newcommand{\up}{\uparrow}

\newcommand{\Di}{\mathit{\Delta}}
\newcommand{\qedm}{\rule{1.5mm}{3mm}}

\newcommand{\ket}[1]{|#1\rangle}
\newcommand{\bra}[1]{\langle#1|}
\newcommand{\La}{\Lambda}
\newcommand{\Htot}{\calH_\mathrm{tot}}
\newcommand{\Dtot}{D_\mathrm{tot}}
\newcommand{\Deff}{D_\mathrm{eff}}
\newcommand{\tDeff}{\widetilde{D}_\mathrm{eff}}
\newcommand{\hP}{\hat{P}}
\newcommand{\Pneq}{\hat{P}^{\Gamma,\ep}_\mathrm{neq}}
\newcommand{\Pneqr}{\hat{P}^{\Gamma,\ep_0(\rho)}_\mathrm{neq}}

\newcommand{\hH}{\hat{H}}
\newcommand{\hU}{\hat{U}}
\newcommand{\hN}{\hat{N}}
\newcommand{\kPj}{\ket{\Psi_j}}
\newcommand{\bPj}{\bra{\Psi_j}}
\newcommand{\kPz}{\ket{\Phi(0)}}
\newcommand{\bPz}{\bra{\Phi(0)}}
\newcommand{\kPt}{\ket{\Phi(t)}}
\newcommand{\bPt}{\bra{\Phi(t)}}
\newcommand{\Tr}{\operatorname{Tr}}

\newcommand{\had}{\hat{a}^\dagger}
\newcommand{\ha}{\hat{a}}
\newcommand{\hbd}{\hat{b}^\dagger}
\newcommand{\hb}{\hat{b}}
\newcommand{\hcd}{\hat{c}^\dagger}
\newcommand{\hc}{\hat{c}}
\newcommand{\hn}{\hat{n}}

\newcommand{\vac}{\ket{\Phi_{\rm vac}}}
\newcommand{\vacb}{\bra{\Phi_{\rm vac}}}

\newcommand{\tn}{\tilde{n}}

\newcommand{\Nel}{N_\mathrm{el}}

\newcommand{\fLh}{\frac{L-1}{2}}
\newcommand{\Lh}{(L-1)/2}
\newcommand{\summu}{\sum_{\mu=1}^{L-1}}
\newcommand{\sumnu}{\sum_{\nu=-\Lh}^{\Lh}}

\newcommand{\bib}{\bibitem}

\renewcommand{\Re}{\operatorname{Re}}
\renewcommand{\Im}{\operatorname{Im}}

\usepackage{color}
\definecolor{fluorescentpink}{rgb}{1.0, 0.08, 0.58}
\definecolor{forestgreen}{rgb}{0.13, 0.55, 0.13}

\begin{document}

\noindent
{\Large\bf Nature abhors a vacuum}

\vspace{1mm}
\noindent{\large\bf
A simple rigorous example of thermalization in an isolated macroscopic quantum system}

\renewcommand{\thefootnote}{\fnsymbol{footnote}}
\medskip\noindent
Naoto Shiraishi\footnote{Faculty of arts and sciences, University of Tokyo, 3-8-1 Komaba, Meguro-ku, Tokyo,
Japan} and Hal Tasaki\footnote{%
Department of Physics, Gakushuin University, Mejiro, Toshima-ku, 
Tokyo 171-8588, Japan.
}
\renewcommand{\thefootnote}{\arabic{footnote}}
\setcounter{footnote}{0}

\begin{quotation}
\small\noindent
We show, without relying on any unproven assumptions, that a low-density free fermion chain exhibits thermalization in the following (restricted) sense.
We choose the initial state as a pure state drawn randomly from the Hilbert space in which all particles are in half of the chain.
This represents a nonequilibrium state such that the half chain containing all particles is in equilibrium at infinite temperature, and the other half chain is a vacuum.
We let the system evolve according to the unitary time evolution determined by the Hamiltonian and, at a sufficiently large typical time, measure the particle number in an arbitrary macroscopic region in the chain.
In this setup, it is proved that the measured number is close to the equilibrium value with probability very close to one.
Our result establishes the presence of thermalization in a concrete model in a mathematically rigorous manner.
The key for the proof is a new strategy to show that a randomly generated nonequilibrium initial state typically has a large enough effective dimension by using only mild verifiable assumptions.
In the present work, we first give general proof of thermalization based on two assumptions, namely, the absence of degeneracy in energy eigenvalues and a property about the particle distribution in energy eigenstates.
We then justify these assumptions in a concrete free-fermion model, where the absence of degeneracy is established by using number-theoretic results.
This means that our general result also applies to any lattice gas models in which the above two assumptions are justified.
To confirm the potential wide applicability of our theory, we discuss some other models for which the essential assumption about the particle distribution is easily verified, and some non-random initial states whose effective dimensions are sufficiently large.
\end{quotation}

\tableofcontents

\section{Introduction}\label{s:introduction}
Whether the unitary time evolution in an isolated macroscopic quantum system can describe the phenomenon of thermalization or, equivalently, the approach to thermal equilibrium is an essential question in the foundation of statistical mechanics.
Since there are several different formulations of thermalization, we shall first make clear what we precisely mean by thermalization in the present work.
Consider a many-body quantum system with Hamiltonian $\hH$ and take a  pure initial state $\kPz$ in which energy is sharply distributed around some value $E$.
We say that the system with this initial state thermalizes if the measurement result of any macroscopic observable $\hat{A}$ in the time-evolved state $\kPt$ after sufficiently long and typical time $t>0$ is indistinguishable (with probability very close to one) from the microcanonical average $\sbkt{\hat{A}}^\mathrm{MC}_E$.
Note that we are dealing with the outcome of a single quantum mechanical measurement of $\hat{A}$ in the state $\kPt$ rather than the quantum mechanical expectation value $\langle\Phi(t)|\hat{A}\kPt$.
Therefore thermalization formulated in this manner guarantees that the result of a single experiment at a sufficiently later time is predicted precisely by equilibrium statistical mechanics.

Our ultimate goal is to rigorously establish the presence of thermalization in the above strong form in a realistic macroscopic quantum system with a realistic nonequilibrium initial state.
But this seems to be a formidably difficult problem for the moment.
In the present paper, we report a partial result toward the goal, namely, complete proof that a low-density free fermion chain exhibits thermalization in the above sense but for a restricted class of observables \cite{video}.

\medskip
The study of thermalization in isolated macroscopic quantum systems goes back to the early days of quantum mechanics \cite{vonNeumann}, but considerable progress has been made in the present century partly motivated by modern ultracold atom experiments~\cite{KWW, Tro, Gri, Lan, Kau}.
It is now a general consensus that a sufficiently complex many-body quantum system has the ability to thermalize only by the unitary time evolution~\cite{DAlessioKafriPolkovnikovRigol2016,GE16}.
\label{refadd2}

An important theoretical concept in the study of thermalization is the energy eigenstate thermalization hypothesis (ETH).
It was first introduced (implicitly) by von Neumann in 1929 \cite{vonNeumann,GLTZ} as an essential assumption for his quantum ergodic theorem.
See \cite{RigolSrednicki2012,DAlessioKafriPolkovnikovRigol2016} for the relation between von Neumann's ETH and the modern version of ETH proposed in \cite{Deutsch1991,Srednicki1994}. 
Another key theoretical concept
is a large effective dimension of the initial state.
It was first pointed out by Tasaki in 1998 \cite{Tasaki1998} (without explicitly introducing the notion of the effective dimension) that one can show the presence of equilibration if the effective dimension is large enough.
It is known that one can prove the presence of thermalization by assuming either (i)~some (strong) form of ETH \cite{vonNeumann,GLTZ,GLMTZ09b,Reimann3,Tasaki2016},  (ii)~some form of ETH and a large enough effective dimension \cite{Tasaki1998,Reimann,LindenPopescuShortWinter,ReimannKastner,Reimann2}, or (iii)~an effective dimension almost as large as the total dimension \cite{GHT14long,Tasaki2016}.

It is strongly believed that the assumptions in the above scenarios (i), (ii), and (iii) are satisfied in a large class of sufficiently complex quantum systems and their realistic (nonequilibrium) initial states.
However, it is extremely difficult, even if not impossible, to justify the assumptions rigorously for concrete models.
As far as we know, there have been no concrete and nontrivial examples of quantum systems with short-range interaction in which the presence of thermalization was justified according to these scenarios without relying on any unproven assumptions.
We note that an example based on a different mechanism is discussed in \cite{Shi17}.

It is interesting, on the other hand, that there have been many examples of quantum systems in which the absence of thermalization was rigorously established.
A well-known example is an integrable system, where the system relaxes not to the equilibrium state but to a state corresponding to an ensemble characterized by its local integrals of motion.
The absence of thermalization in such systems with local integrals of motion is an old established property~\cite{Grad, Jancel}, and has recently been studied in detail in terms of the generalized Gibbs ensemble~\cite{Caz, RDYO}.
Another example is a system with many-body localization:
A spin system with many-body localization has random interactions or a random magnetic field, and this randomness prohibits its thermalization as in the case of the Anderson localization~\cite{BAA, PH, SPA, Fri15, NH15, Imb}.
Recently, a more exotic system was found where most initial states thermalize while some do not.
This phenomenon was first observed in experiments of cold atoms~\cite{Ber}, and independently from this experiment, a general theoretical framework covering such phenomena was proposed~\cite{SM, MS}.
Later, such phenomena were named quantum many-body scar states, and have attracted the interests of broad research fields~\cite{Tur18, Ber17, Mou, LM18, Ho, Shi19, SAP}.
Furthermore, it has even been shown that the problem of thermalization is, in general, undecidable \cite{ShiraishiMatsumoto}.

\medskip
The goal of this paper is to present a nontrivial and rigorous concrete example of thermalization (in a restricted sense) that does not rely on any unproven assumptions.
We first develop a general theory of unitary time evolution in a low-density lattice gas that satisfies two crucial assumptions, and establish the presence of thermalization with respect to the number operator for any macroscopic region, assuming that the initial nonequilibrium state is generated randomly.
The derivation is based on the above-mentioned scenario (iii), which requires the effective dimension of the initial state to be almost as large as the total Hilbert space dimension.
We then prove that the two assumptions are indeed satisfied in the simplest model, namely, the free fermion chain with suitable parameters.
Although a free fermion system does not exhibit full-fledged thermalization, i.e., the approach to thermal equilibrium from an arbitrary nonequilibrium state (with almost fixed energy), it does thermalize in our setting where the initial state is sufficiently complex.
See \cite{RigolMuramatsuOlshanii2006,RigolFitzpatrick2011,Pandeyetal} for detailed numerical studies of closely related problems.
\label{addref1}

It is important to note, however, that our general theory should apply to non-integrable models as well, in which one expects full-fledged thermalization to take place.
In fact, the key assumption in our theory is about the particle distribution in energy eigenstates, which may be regarded as a very restrictive form of ETH.
The other assumption is the absence of degeneracy in the energy spectrum of the model, which appears highly natural and plausible in complex many-body systems.
Interestingly, if we assume the absence of degeneracy, we can justify the first assumption about the particle distribution for a wider class of lattice gas models, including interacting ones.
It is an intriguing problem whether one can find non-integrable models in which our assumptions can be fully justified.

\medskip
Before going into details of our theory, let us state precisely what we can prove for free fermion chains.
Consider a system of $N$ fermions on the chain $\{1,\ldots,L\}$, where we fix the density $\rho=N/L$ and make $N$ and $L$ large.
We take the standard Hamiltonian with uniform nearest-neighbor hopping
\eq
\hH=\sum_{x=1}^L\bigl\{e^{i\theta}\,\hcd_x\hc_{x+1}+e^{-i\theta}\,\hcd_{x+1}\hc_x\bigr\},
\lb{H0}
\en
where the phase $\theta\in\bbR$ is introduced (artificially) to break the reflection symmetry. 
See section~\ref{s:fermionEE} for notations and details.
In the most standard model with $\theta=0$, most energy eigenvalues are degenerate because of the reflection symmetry (which brings the wave number $k$ to $-k$).
It is likely that the degeneracies are lifted by a nonzero phase $\theta$.
We assume that the parameters are properly chosen so that all the energy eigenvalues of $\hH$ are nondegenerate.
In fact, we prove in section~\ref{s:nondeg} that the model is free from degeneracy under some conditions.
For example, it suffices to set $\theta=(4N+2L)^{-\Lh}$ provided that $L$ is an odd prime.

We choose initial state $\kPz$ randomly from the subspace of states in which all fermions are in the half-chain $\{1,\ldots,(L-1)/2\}$.
This corresponds to the infinite temperature equilibrium state confined in the half-chain.
We then denote by $\kPt=e^{-i\hH t}\kPz$ the state at time $t>0$.
We let $\hN_\mathrm{left}$ be the operator that counts the number of fermions on the half-chain $\{1,\ldots,(L-1)/2\}$.
Then, our main result is as follows:
\begin{theorem}\label{t:main0}
When $N$ (or $L$) is sufficiently large and $\rho=N/L\le1/5$, the following is true with probability larger than $1-e^{-(\rho/3)N}$ (where the probability is that for the choice of $\kPz$).
There exists a sufficiently long time $T>0$ and a subset (a collection of intervals) $G\subset[0,T]$ with $\mu(G)/T\ge1-e^{-(\rho/4)N}$ (where $\mu(G)$ denotes the total length of the intervals in $G$) such that if one measures $\hN_\mathrm{left}$ in $\kPt=e^{-i\hH t}\kPz$ at any $t\in G$ the measurement result $N_\mathrm{left}$ satisfies 
\eq
\Bigl|\frac{N_\mathrm{left}}{N}-\frac{1}{2}\Bigr|\le\ep_0(\rho),
\lb{NG12}
\en
with probability larger than $1-e^{-(\rho/4)N}$ (where the probability is that for quantum measurement).
Here we set $\ep_0(\rho)=\sqrt{\frac{3}{2}\rho}$.
\end{theorem}
The factors $e^{-(\rho/3)N}$ and $e^{-(\rho/4)N}$ are essentially negligible if $\rho N\gg1$.
Then the theorem states that it almost certainly happens that the measurement result of $\hN_\mathrm{left}/N$ at a sufficiently large and typical time is close to its equilibrium value $1/2$ with precision $\ep_0(\rho)$.
Since the measurement result of $\hN_\mathrm{left}/N$ is 1 in the initial state $\kPz$, this establishes an irreversible behavior (or 
the approach to thermal equilibrium) with respect to the observable $\hN_\mathrm{left}$.
We should note that our result is not limited for a single specific observable $N_{\rm left}$.
In fact, the main theorem, Theorem~\ref{t:main}, is stated for the number operator for any macroscopic region.
As we have already stressed, it is crucial that we are dealing with the result of a single projective measurement of $\hN_\mathrm{left}$ in the state $\kPt$, rather than its quantum mechanical average $\langle\Phi(t)|\hN_\mathrm{left}\kPt$.

We must note, however, that the precision $\ep_0(\rho)$ in \rlb{NG12} is a function of the density $\rho$, and may not be small.
One needs to consider a system with low density in order to have high precision.
For example, $\rho\simeq10^{-4}$ for $\ep_0\simeq10^{-2}$, or $\rho\simeq0.04$ for $\ep_0=1/4$.
This density-dependence of the precision $\ep_0(\rho)$ is a major shortcoming of the present theory, which reflects our strategy to base the theory only on mild verifiable assumptions.
 We nevertheless stress that our theorem establishes thermalization in a certain sense without relying on any unproven assumptions.
 
 \medskip
 The present paper is organized as follows.
 In section~\ref{s:main}, we state our main thermalization theorem for a general lattice gas satisfying two assumptions, namely, Assumptions~\ref{a:nondeg} and \ref{a:P2N}.
Then in section~\ref{s:freefermion}, we prove these two assumptions are indeed satisfied in free fermion chains with suitable parameters.

In Appendix~\ref{s:deg}, we discuss the extension of our general theory to a model in which the energy spectrum is moderately degenerate.
In Appendix~\ref{s:doublelattice}, we present two classes of models (one of which includes non-integrable models) in which we can justify Assumption~\ref{a:P2N} about the particle distribution in energy eigenstates, assuming that the energy eigenvalues are nondegenerate.
We stress that Assumption~\ref{a:P2N} is indeed an essential nontrivial assumption in our theory.
In Appendix~\ref{s:Deffx}, we present some concrete estimates of the effective dimensions of some non-random initial states in the free fermion chain, and with the help of this estimate, we prove that some non-random initial states indeed thermalize.
Finally, in Appendix~\ref{s:FiniteT}, we briefly discuss a possible extension of our result to finite temperature states.

\section{General results}
\label{s:main}
Here we describe general systems of lattice gas, state necessary assumptions, and prove the main low-density thermalization theorem.
The new observation about the effective dimension is summarized in Theorem~\ref{t:Deff}.

\subsection{Setting and main assumptions}
\label{s:setting}
Let $\La$ be a lattice with $L$ sites, and consider a system of $N$ fermions on $\La$.\footnote{All the results in section~\ref{s:main} are also valid for a system of hardcore bosons.}
A typical example is the chain $\La=\{1,\ldots,L\}$.
We take the thermodynamic convention (except in Appendix~\ref{s:Deffx}), in which we fix the density $\rho$, choose $L$ and $N$ such that $N/L\simeq\rho$, and make $L$ and $N$ sufficiently large.
Our results are meaningful in the low-density regime, where $\rho$ is sufficiently small.

Let $\Htot$ be the Hilbert space of the system with $N$ particles on the lattice $\La$.
The dimension $\Dtot$ of $\Htot$ is given by
\eq
\Dtot=\binom{L}{N}\sim e^{L\,S(\rho)},
\lb{Dtot}
\en
where the relation $F(L)\sim G(L)$ means
\eq
\lim_{L\up\infty}\frac{1}{L}\log\frac{F(L)}{G(L)}=0,
\en
and
\eq
S(p)=-p\log p-(1-p)\log(1-p),
\lb{Srho}
\en
is the binominal entropy.
The final expression in \rlb{Dtot} comes from the Stirling formula.

We decompose the lattice $\La$ disjointly into two parts as $\La=\La_1\cup\La_2$, where $|\La_1|=(L-1)/2$ and $|\La_2|=(L+1)/2$ when $L$ is odd, and $|\La_1|=|\La_2|=L/2$ when $L$ is even.
Throughout the present paper, we denote by $|S|$ the number of elements in a set $S$.
Let $\calH_1$ denote the nonequilibrium subspace where all particles are in the sublattice $\La_1$ and hence $\La_2$ is empty.
The dimension of $\calH_1$ is
\eq
D_1=\binom{\frac{L-1}{2}}{N}\sim e^{(L/2)S(2\rho)},
\lb{D0}
\en
where we assumed $L$ is odd (but the result is essentially the same for even $L$).
We denote by $\hP_1$ the projection onto the subspace $\calH_1$.

Let $\hH$ be the Hamiltonian of the system.
We assume that $\hH$ preserves the particle number, and denote by $\kPj\in\Htot$ with $j=1,\ldots,\Dtot$ its normalized eigenstate (with $N$ particles) corresponding to the energy eigenvalue $E_j$.
We make two essential assumptions about energy eigenvalues and eigenstates.
\begin{assumption}\label{a:nondeg}
The energy eigenvalues $E_1,\ldots,E_{\Dtot}$ of $\hH$ are nondegenerate.
\end{assumption}

It is believed that the energy eigenvalues of a quantum many-body system are, in general, nondegenerate unless there are special reasons (such as symmetry) that cause degeneracy.
In other words, it is likely that accidental degeneracies can be always lifted by adding an appropriate small perturbation to the Hamiltonian.
It is however not at all easy to make this intuition into proof for a concrete class of models.
In section~\ref{s:nondeg}, we shall prove that some free fermion models on a chain are indeed free from degeneracy.
See Theorems~\ref{T:nondeg} and \ref{T:nondeg2}.
\begin{assumption}\label{a:P2N}
For any $j=1,\ldots,\Dtot$, the energy eigenstate $\kPj$ satisfies
\eq
\bPj\hP_1\kPj\le2^{-N}.
\lb{2N}
\en
\end{assumption}

Here $\bPj\hP_1\kPj$ is the probability to find all particles in $\La_1$ in the state $\kPj$.
Note that one gets the probability $2^{-N}$ if each particle independently chooses between $\La_1$ and $\La_2$ with probability 1/2.
The bound \rlb{2N} is reasonable since the hardcore nature further reduces the probability.
We expect the bound \rlb{2N} to hold for a large class of interacting quantum lattice gases, but, for the moment, we are able to prove it for a class of non-interacting fermions (section~\ref{s:P2N} and Appendix~\ref{s:A2}) and systems of interacting fermions or hardcore bosons on a double lattice with special symmetry (Appendix~\ref{s:A1}).

We also note that Assumption~\ref{a:P2N} is reminiscent of the strong ETH in the sense that it is an assertion about every energy eigenstate.
But this is much weaker than the standard ETH since we only require that a single observable, rather than all macroscopic observables, satisfies the inequality \rlb{2N}, rather than an equality.

\medskip

In what follows, we first show that, under Assumption~\ref{a:P2N}, a random initial state has an extremely large effective dimension with high probability (Theorem~\ref{t:Deff}).
Then, by combining Assumption~\ref{a:nondeg} and the largeness of effective dimension, we conclude that this initial state thermalizes (Theorem~\ref{t:main}).

\subsection{Initinal state and its effective dimension}
\label{s:Deff}
Let $\kPz\in\Htot$ be the normalized initial state of the system.
We define the effective dimension $\Deff$ of $\kPz$ by
\eq
\Deff=\biggl(\sum_{j=1}^{\Dtot}\bigl|\bPz\Psi_j\rangle\bigr|^4\biggr)^{-1},
\lb{Deff}
\en
which quantifies the effective number of energy eigenstates that constitute the state $\kPz$.
It holds in general that $1\le\Deff\le\Dtot$.
It is known that an initial state whose effective dimension $\Deff$ is almost as large as $\Dtot$ generically exhibits thermalization, provided that the energy eigenvalues are nondegenerate.
See section~6 of  \cite{Tasaki2016}.
(See Appendix~\ref{s:deg} for necessary modifications when there are degeneracies.)
It is strongly believed that a realistic nonequilibrium initial state of a non-integrable many-body quantum system has an effective dimension almost as large as the total Hilbert space dimension\footnote{%
\label{f:Deff}
To be precise this is true only when the final state represents the equilibrium state at infinite temperature (as in our case).
In general, if the initial state $\kPz$ has energy close to $E$ then the effective dimension is believed to be close to the dimension of the corresponding energy shell, i.e., the Hilbert space spanned by energy eigenstates whose eigenvalues are close to $E$.

One can argue, although very heuristically, that a large effective dimension is a consequence of (a strong form of) ETH.
Consider a system described by a short-ranged translation-invariant Hamiltonian $\hH$ and assume that ETH is valid.
For simplicity, we take the initial state $\kPz$ to be a translation invariant product state.
(We assume $\kPz$ is not an eigenstate of $\hH$.)
Then $\kPz$ has energy distribution peaked around some value $E$.
Let $\kPj$ be the eigenstate of $\hH$ with eigenvalue $E_j$.
Since ETH asserts that energy eigenstates with close eigenvalues are similar to each other, it is reasonable to assume that the overlap $|\langle\Phi(0)\kPj|^2$ is almost independent of $j$ as long as $E_j\simeq E$.
This implies that $\Deff$ is almost identical to the dimension of the energy shell around $E$.
}.
See \cite{SantosPolkovnikovRigol2011,Rigol2014,Rigol2016} for systematic convincing numerical studies.\footnote{\label{fn:Sd}
We note that the diagonal entropy $S_{\rm d}$ studied in these works is believed to be related to the effective dimension as $\Deff\sim\exp[S_{\rm d}]$.}
However, it seems to be formidably difficult to prove this expectation rigorously.
Currently available general lower bound for $\Deff$ only shows that it is only moderately large \cite{FBC2017}.
Our major task is to construct an initial state $\kPz$ that is far from equilibrium and has a large effective dimension $\Deff$.

To realize such an initial state with large $\Deff$, we choose $\kPz$ randomly from the subspace $\calH_1$.
To be precise, denoting by $\{\ket{\Xi_j}\}_{j=1,\ldots,D_1}$ an arbitrary orthonormal basis of $\calH_1$ we prepare an initial state as $\kPz=\sum_{j=1}^{D_1}c_j\ket{\Xi_j}$, where $c_j\in\bbC$ satisfies $\sum_j|c_j|^2=1$ and are drawn randomly according to the uniform measure on the unit sphere in the $D_1$ dimensional complex space.
Such $\kPz$ describes a nonequilibrium state such that all particles are confined in the sublattice $\La_1$, while the state restricted to $\La_1$ is in thermal equilibrium at infinite temperature.
In this state, the infinite temperature state in $\La_1$ borders a vacuum in $\La_2$. 
Therefore we can interpret the present initial state as a limiting case of a nonequilibrium state in which two equilibrium states with different pressures are in touch with each other.

We then have the following essential result, which is the main new observation in the present paper.
\begin{theorem}\label{t:Deff}
Suppose that Assumption~\ref{a:P2N} is valid and that $\rho\le1/5$.
Then, for sufficiently large $N$, one has
\eq
\frac{\Dtot}{\Deff}\le e^{\rho N},
\lb{DDbound}
\en
with probability larger than $1-e^{-(\rho/3)N}$.
\end{theorem}

Here the probability is that for the random choice of the initial state $\kPz$.
We thus see that, when $\rho$ is small, the effective dimension $\Deff$ is almost as large as $\Dtot$ with probability very close to one.
We shall see in section~\ref{s:TET} below that the upper bound \rlb{DDbound} implies thermalization in a certain sense.

\medskip
\noindent
{\em Proof of Theorem~\ref{t:Deff}:}\/
It is well known (and can easily be shown) that for any $\ket{\Xi}\in\calH_1$, one has
\eq
\overline{\bigl|\bPz\Xi\rangle\bigr|^4}=\frac{2}{D_1(D_1+1)}\,\snorm{\ket{\Xi}}^4,
\lb{DD}
\en
where the bar on the left-hand side denotes the average over the random choice of $\kPz$.
See, e.g., \cite{Ullah}. 
Noting that $\bPz\Psi_j\rangle=\bPz\hP_1\kPj$ and that $\hP_1\kPj\in\calH_1$, we find from \rlb{Deff} and \rlb{DD} that
\eq
\overline{\Deff^{-1}}
=\sum_{j=1}^{\Dtot}\overline{\bigl|\bPz\hP_1\kPj\bigr|^4}
=\frac{2}{D_1(D_1+1)}\sum_{j=1}^{\Dtot}\snorm{\hP_1\kPj}^4.
\lb{Deff2}
\en
By using the assumed bound \rlb{2N}, which is written as $\snorm{\hP_1\kPj}^2\le2^{-N}$, we find
\eqa
\overline{\Deff^{-1}}&\le\frac{2}{D_1(D_1+1)2^N}\sum_{j=1}^{\Dtot}\snorm{\hP_1\kPj}^2
=\frac{2}{D_1(D_1+1)2^N}\Tr[\hP_1]
=\frac{2}{(D_1+1)2^N},
\lb{Deff23}
\ena
where we noted that $\Tr[\hP_1]=D_1$.
Recalling \rlb{Dtot} and \rlb{D0}, we see that
\eq
\Dtot\overline{\Deff^{-1}}\le\frac{2\Dtot}{2^ND_1}\sim \exp[L\,S(\rho)-\tfrac{L}{2}S(2\rho)-N\log2]=e^{g(\rho)L},
\lb{DD2}
\en
with
\eqa
g(\rho)&=S(\rho)-\tfrac{1}{2}S(2\rho)-\rho\log2
\nl&=-(1-\rho)\log(1-\rho)+\frac{1-2\rho}{2}\log(1-2\rho)
\nl&=\frac{\rho^2}{2}+\frac{\rho^3}{2}+\frac{7\rho^4}{12}+\cdots<\frac{2}{3}\rho^2.
\ena
Here the final inequality is verified for $\rho\in[0,1/5]$ with an aid of numerical evaluation.
We can rewrite the estimate \rlb{DD2} into the bound
\eq
\Dtot\overline{\Deff^{-1}}\le\exp[\tfrac{2}{3}\rho^2L]=\exp[\tfrac{2}{3}\rho N],
\lb{DtDe}
\en
provided that $L$ (or $N$) is sufficiently large.
Theorem~\ref{t:Deff} then follows from Markov's inequality as follows.
Let $p$ be the probability that $\Dtot\Deff^{-1}$ is larger than $e^{\rho N}$.
Then we see $\Dtot\overline{\Deff^{-1}}\ge p\,e^{\rho N}$, which, with \rlb{DtDe}, implies $p\le e^{-(\rho/3)N}$.~\qedm

\medskip
One may prefer a statement for a definite (i.e., non-random) initial state rather than that for (the majority of) random initial states.
In Appendix~\ref{s:Deffx}, we discuss a non-random initial state whose effective dimension almost saturates as in Theorem~\ref{t:Deff}.

\subsection{Time evolution and thermalization}
\label{s:TET}
Let us now consider the state obtained from the initial state $\kPz$ by the unitary time evolution, i.e.,
\eq
\kPt=e^{-i\hH t}\kPz=\sum_{j=1}^{\Dtot}e^{-iE_jt}\kPj\sbkt{\Psi_j|\Phi(0)}.
\lb{Phit}
\en
We expect that, for sufficiently large and typical $t$, the time-evolved state $\kPt$ describes (in a certain physical sense) the thermal equilibrium at infinite temperature.
See the next subsection.

To examine the property of the state $\kPt$, we take an arbitrary subset $\Gamma$ of $\La$ such that $|\Gamma|=\gamma L$, where $\gamma$ is a constant of order 1, and measure the proportion of particles in $\Gamma$.
We shall prove that, for sufficiently large and typical time $t$, the proportion is close to its equilibrium value, $\gamma$, with probability very close to one.
This type of statement has been shown in the literature for initial states with extremely large effective dimensions~\cite {GHT14long,Tasaki2016}, and we follow the standard idea. 
Our precise statement is as follows.
\begin{theorem}\label{t:main}
We fix the (small) density $\rho>0$, and take sufficiently large $L$ and $N$ such that $N/L\simeq\rho$.
We consider a system of $N$ particles on the lattice $\La$ such that $|\La|=L$ and let $\hH$ be the Hamiltonian.
Suppose that Assumption~\ref{a:nondeg} about nondegeneracy is valid and also that the effective dimension $\Deff$ is large enough to satisfy the bound \rlb{DDbound}.  (This is guaranteed by Theorem~\ref{t:Deff} to be extremely likely.)
Take any $\Gamma\subset\La$ such that $|\Gamma|=\gamma L$, and let $\hN_\Gamma$ be the operator that counts the number of particles in $\Gamma$.
Then there exists a constant $T>0$ and a subset (a collection of intervals) $G\subset[0,T]$ with
\eq
\frac{\mu(G)}{T}\ge 1-e^{-(\rho/4)N},
\lb{TG}
\en
where $\mu(G)$ is the total length of the intervals in $G$.
Suppose that one performs a measurement of the number operator $\hN_\Gamma$ in the state $\kPt$ with arbitrary $t\in G$.
Then, with probability larger than $1-e^{-(\rho/4)N}$, the measurement result $N_\Gamma$ satisfies
\eq
\Bigl|\frac{N_\Gamma}{N}-\gamma\Bigr|\le\ep_0(\rho),
\lb{NGg}
\en
where the precision is given by
\eq
\ep_0(\rho)=\sqrt{6\gamma(1-\gamma)\rho}.
\lb{ep0}
\en
\end{theorem}

Here the probability is that for the quantum mechanical measurement.
Suppose that $N$ is sufficiently large so that $e^{-(\rho/4)N}$ is negligibly small.
Then the theorem guarantees that \rlb{NGg} almost certainly holds for almost all $t$ in $[0,T]$.
The bound \rlb{NGg} states that the observed proportion $N_\Gamma/N$ is close to its equilibrium value, $\gamma$.
Recalling that the initial state $\kPz$ is a nonequilibrium state in which all particles are in $\La_1$, we have established that the system thermalizes only by means of unitary time evolution \rlb{Phit}.

We must note, however, that the precision in the relation \rlb{NGg} is given by $\ep_0(\rho)$, which is a function of $\rho$ as in \rlb{ep0} and may not be small.
In fact, we need to make the density $\rho$ sufficiently low to achieve high precision.
If one demands that the precision $\ep_0(\rho)$ should be, for example, of order $10^{-2}$ then $\rho$ should be of order $10^{-4}$.
This density dependence of the precision and the resulting limitation to dilute gases are the major shortcomings of the present theory, which comes from our strategy to base the whole theory on mild verifiable assumptions, namely, Assumptions~\ref{a:nondeg} and \ref{a:P2N}.

We should also remark that our criterion for thermal equilibrium deals only with the number of particles in an arbitrary macroscopic region.
We have proved the presence of thermalization, but only with respect to this rather restricted criterion.
This again reflects the limitation arising from our mild assumptions.
Although we expect that thermalization for other macroscopic quantities reflecting single-particle properties can be established by a straightforward extension of the present analysis, we are far from treating quantities that involve particle-particle correlations.
See the discussion at the end of section~\ref{s:discussion}.

\medskip
\noindent
{\em Proof of Theorem~\ref{t:main}:}\/
The proof consists essentially of a combination of standard arguments found in the literature.
For $\ep>0$, let $\Pneq$ denote the projection operator onto the subspace of $\Htot$ determined by
\eq
\biggl|\frac{\hN_\Gamma}{N}-\gamma\biggr|\ge\ep.
\lb{NGgamma}
\en
Clearly, the expectation value $\bPt\Pneqr\kPt$ is the probability that the measurement result of $\hN_\Gamma$ in $\kPt$ does not satisfy the relation \rlb{NGg}.
From \rlb{Phit}, we see that
\eq
\bPt\Pneq\kPt=\sum_{j,j'=1}^{\Dtot}e^{i(E_j-E_{j'})t}
\sbkt{\Phi(0)|\Psi_{j}}\bPj\Pneq\ket{\Psi_{j'}}\sbkt{\Psi_{j'}|\Phi(0)}.
\en
Since we assumed that the energy eigenvalues $E_j$ are non-degegerate, the long-time average of $\bPt\Pneq\kPt$ is expressed in terms of a single sum as
\eqa
\lim_{T\up\infty}\frac{1}{T}\int_0^Tdt\,\bPt\Pneq\kPt
&=\sum_{j=1}^{\Dtot}\bigl|\sbkt{\Phi(0)|\Psi_{j}}\bigr|^2\bPj\Pneq\kPj
\nl&\le\sqrt{\biggl(\sum_{j=1}^{\Dtot}\bigl|\sbkt{\Phi(0)|\Psi_{j}}\bigr|^4\biggr)
\biggl(\sum_{j=1}^{\Dtot}\bPj\Pneq\kPj^2\biggr)
}
\nl&\le\sqrt{
\Dtot\Deff^{-1}\,\sbkt{\Pneq}_\infty,
}
\lb{PPP1}
\ena
where we  defined the canonical average at infinite temperature by 
\eq
\sbkt{\cdots}_\infty=\frac{\Tr_{\Htot}[\cdots]}{\Dtot}.
\en
In \rlb{PPP1}, the second line follows from the Schwarz inequality, and the final expression follows from \rlb{Deff} by noting $\bPj\Pneq\kPj^2\le\bPj\Pneq\kPj$.

Below we prove the large-deviation type upper bound
\eq
\sbkt{\Pneq}_\infty\le C\exp\Bigl[-\frac{\ep^2}{3\gamma(1-\gamma)}N\Bigr]
=C\exp\Bigl[-2\rho\Bigl(\frac{\ep}{\ep_0(\rho)}\Bigr)^2N\Bigr],
\lb{PC}
\en
with a constant $C>1$, assuming that $N$ is sufficiently large and $\ep$ is sufficiently small.
Note that the right-hand side reduces to $Ce^{-2\rho N}$ if we set $\ep=\ep_0(\rho)$.
Recalling  \rlb{DDbound}, we find that the right-hand side of  \rlb{PPP1} with $\ep=\ep_0(\rho)$ is bounded from above by $\sqrt{C}\,e^{-(\rho/2)N}$.
This means that there is sufficiently large $T$ such that the finite-time average satisfies
\eq
\frac{1}{T}\int_0^Tdt\,\bPt\Pneqr\kPt\le e^{-(\rho/2)N}.
\lb{PPP2}
\en

To rewrite the obtained relation into the form of Theorem~\ref{t:main}, we apply Markov's inequality.
We let $G$ be the set of time at which \rlb{NGg} is satisfied with probability larger than $1-e^{-(\rho/4)N}$:
\eq
G=\bigl\{t\in[0,T]\,\bigl|\,\bPt\Pneqr\kPt\le e^{-(\rho/4)N}\bigr\}.
\lb{Gdef}
\en
The property of $G$ stated in the theorem is fulfilled by construction.
It remains to verify \rlb{TG} for the above $G$.
For this, it suffices to note that
\eq
\frac{1}{T}\int_0^Tdt\,\bPt\Pneqr\kPt\ge\frac{1}{T}\int_{t\in[0,T]\backslash G}dt\,e^{-(\rho/4)N}
=\frac{T-\mu(G)}{T}\,e^{-(\rho/4)N},
\en
which, with \rlb{PPP2}, implies the desired \rlb{TG}.~\qedm

\medskip
\noindent
{\em Derivation of \rlb{PC}:}\/
We shall be brief since the derivation is standard and elementary.
Let $\hP_M$ be the projection onto the subspace with $\hN_\Gamma=M$.
It is clear that
\eq
\Pneq=\sumtwo{M}{(|M/N-\gamma|\ge\ep)}\hP_M,
\en
and
\eq
\sbkt{\hP_M}_\infty\sim\exp\biggl[\gamma L\,S\Bigl(\frac{M}{\gamma L}\Bigr)+(1-\gamma)L\,S\Bigl(\frac{N-M}{(1-\gamma)L}\Bigr)-L\,S\Bigl(\frac{N}{L}\Bigr)\biggr].
\en
When $|M/N-\gamma|=\ep$ or, equivalently, $M/N-\gamma=\pm\ep$, the two first argument of $S(\cdot)$ in the above expression read
\eq
\frac{M}{\gamma L}=\Bigl(1\pm\frac{\ep}{\gamma}\Bigr)\rho,\quad
\frac{N-M}{(1-\gamma)L}=\Bigl(1\mp\frac{\ep}{1-\gamma}\Bigr)\rho.
\en
Since $\sbkt{\hP_M}_\infty$ takes a very sharp maximum around $M$ such that $M/(\gamma L)=\rho$, the probability that $|M/N-\gamma|\ge\ep$ is almost the same as the probability that $|M/N-\gamma|\simeq\ep$.
We thus have
\eqa
\sbkt{\Pneq}_\infty&\sim\max_\pm\exp\bigl[\gamma L\,S\bigl((1\pm\tfrac{\ep}{\gamma})\rho\bigr)+(1-\gamma)L\,S\bigl((1\mp\tfrac{\ep}{1-\gamma})\rho\bigr)-L\,S(\rho)\bigr]
\nl&=\exp\biggl[-\Bigl\{\frac{1}{2}\frac{1}{\gamma(1-\gamma)}\frac{\rho}{1-\rho}\ep^2+O(\ep^3)\Bigr\}\,L\biggr].
\ena
For sufficiently large $L$ and small $\ep$ this is converted into the inequality \rlb{PC}.~\qedm

\subsection{Nature of the final state}
\label{s:Natureoffinalstates}
As we have noted several times, we expect that the state $\kPt$ with sufficiently large and typical $t$ represents (with certain limited accuracy) the thermal equilibrium state of the whole system at infinite temperature.
Here we briefly explain why the infinite temperature state, rather than a finite temperature state, is the destination of the relaxation process.

Let us assume in general that the Hamiltonian is written as
\eq
\hH=\hH_1+\hH_2+\Di\hH,
\lb{Hdec}
\en
where $\hH_1$ and $\hH_2$ act only on $\La_1$ and $\La_2$, respectively, and $\Di\hH$ is the interaction Hamiltonian between $\La_1$ and $\La_2$.
We assume that $\hH_1$ and $\hH_2$ are almost identical and $\Di\hH$ is smaller.\footnote{$\Di\hH$ may not be small in the class of models considered in Appendix~\ref{s:doublelattice}.}
We shall use the standard convention that $\hH_1\vac=\hH_2\vac=\Di\hH\vac=0$, where $\vac$ is the state with no particles.
Then we see from the energy conservation that
\eq
\bra{\Phi(t)}\hH\kPt=\bra{\Phi(0)}\hH\kPz\simeq\bra{\Phi(0)}\hH_1\kPz\simeq\frac{\Tr_{\calH_1}[\hH_1]}{\Tr_{\calH_1}[\hat{1}]},
\en
where we recalled that $\kPz$ is drawn randomly from $\calH_1$.
In a standard lattice gas model at low density, we expect from extensivity that
\eq
\frac{\Tr_{\calH_1}[\hH_1]}{\Tr_{\calH_1}[\hat{1}]}\simeq\frac{\Tr_{\Htot}[\hH]}{\Tr_{\Htot}[\hat{1}]}\simeq N\epsilon_\infty,
\en
where $\epsilon_\infty$ is the energy per particle in the equilibrium state at infinite temperature.
We thus see 
\eq
\bra{\Phi(t)}\hH\kPt\simeq N\epsilon_\infty,
\en 
i.e., $\kPt$ has almost the same energy as the equilibrium state of the whole system at infinite temperature.
This is confirmed explicitly for the free fermion chain.
In summary, if the initial state $\kPz$ has an almost saturating effective dimension, then the state after time evolution $\kPt$ represents thermal equilibrium at infinite temperature.

\section{Free fermion on the chain}
\label{s:freefermion}
In this section, we discuss our concrete example, namely the one-dimensional system of free fermions.
We shall show that the model satisfies Assumptions~\ref{a:nondeg} and \ref{a:P2N} if we take a suitable setting.

\subsection{Energy eigenstates and eigenvalues}\label{s:fermionEE}

We consider the chain $\La=\{1,2,\ldots,L\}$, where $L$ is odd.
We denote the sites as $x,y,\ldots\in\La$.
Let $\hc_x$ and $\hcd_x$ be the annihilation and creation operators, respectively, of the fermion at site $x\in\La$.
They satisfy the canonical anticommutation relations $\{\hc_x,\hc_y\}=0$ and $\{\hc_x,\hcd_y\}=\delta_{x,y}$ for any $x,y\in\La$, where $\{\hat{A},\hat{B}\}=\hat{A}\hat{B}+\hat{B}\hat{A}$.
We denote by $\vac$ the state with no particles.

We take the standard Hamiltonian
\eq
\hH=\sum_{x=1}^L\bigl\{e^{i\theta}\,\hcd_x\hc_{x+1}+e^{-i\theta}\,\hcd_{x+1}\hc_x\bigr\},
\lb{H}
\en
where we set the hopping amplitude to be unity for convenience.
We introduced the artificial phase factor $\theta\in[0,2\pi)$ in order to avoid degeneracy.
We impose the periodic boundary condition and identify $\hc_{L+1}$ with $\hc_1$.

The Hamiltonian $\hH$ is readily diagonalized in terms of the plane wave states.
Setting the $k$-space as
\eq
\calK=\Bigl\{\frac{2\pi}{L}\nu\,\Bigl|\,\nu=0,\pm1,\ldots,\pm\fLh\Bigr\},
\lb{K}
\en
we define the creation operator
\eq
\had_k=\frac{1}{\sqrt{L}}\sum_{x=1}^Le^{ikx}\,\hcd_x,
\lb{ak}
\en
for $k\in\calK$.
It holds that $\{\had_k,\ha_{k'}\}=\delta_{k,k'}$.
One can show from the basic anticommutation relations that 
\eq
[\hH,\had_k]=2\tau\cos(k+\theta)\,\had_k.
\lb{Hak}
\en
Let $\bsk=(k_1,\ldots,k_N)$ denote a collection of $N$ elements in $\calK$ such that $k_j<k_{j+1}$ for $j=1,\ldots,N-1$, and define
\eq
\ket{\Psi_{\bsk}}=\had_{k_1}\had_{k_2}\cdots\had_{k_N}\vac.
\lb{Psik}
\en
From \rlb{Hak} we readily see that $\ket{\Psi_{\bsk}}$ is an energy eigenstate, i.e.,
\eq
\hH\ket{\Psi_{\bsk}}=E_{\bsk}\ket{\Psi_{\bsk}},
\en
where the energy eigenvalue is 
\eq
E_{\bsk}=2\sum_{j=1}^N\cos(k_j+\theta).
\lb{Ek}
\en
By counting the dimension, we see that these are the only energy eigenstates and eigenvalues.

\subsection{Justification of Assumption~\ref{a:nondeg}}
\label{s:nondeg}
We prove two theorems for the free fermion chain that justify Assumption~\ref{a:nondeg} about the absence of degeneracy in the energy eigenvalues.
Note that the free fermion model on the continuous interval always has degenerate many-body energy eigenvalues.
The degeneracy cannot be lifted by the flux insertion (which corresponds to the phase $\theta$).
The following results on nondegeneracy essentially rely on the lattice nature of the model.

The first theorem rules out the degeneracy for most values of $\theta$.\footnote{
The theorem was proved by one of us in \cite{Tasaki2010}.
See also Proposition~10.1 in \cite{Tasaki2016} for a similar statement for a slightly complicated model.
}
\begin{theorem}[Nondegeneracy of $E_{\bsk}$ for most $\theta$]\label{T:nondeg}
Let $L$ be an arbitrary odd prime and $N$ be an arbitrary integer with $0<N\le L$.
Except for a finite number of $\theta\in[0,2\pi)$, one has $E_{\bsk}\ne E_{\bsk'}$ whenever $\bsk\ne\bsk'$, i.e., the energy eigenvalues $E_{\bsk}$ are nondegenerate.
\end{theorem}
The theorem, in particular, implies that if one draws $\theta$ randomly, then with probability one, all the energy eigenvalues $E_{\bsk}$ are nondegenerate.
The second theorem allows one to choose a model free from degeneracy without relying on a probabilistic choice.
\begin{theorem}[Nondegeneracy of $E_{\bsk}$ for small $|\theta|\ne0$]\label{T:nondeg2}
Let $L$ be an arbitrary odd prime and $N$ be an arbitrary integer with $0<N\le L$.
For any $\theta\ne0$ such that
\eq
|\theta|\le\frac{1}{(4N+2L)^{\Lh}},
\lb{thetacond}
\en
one has $E_{\bsk}\ne E_{\bsk'}$ whenever $\bsk\ne\bsk'$, i.e., the energy eigenvalues $E_{\bsk}$ are nondegenerate.
\end{theorem}
One thus knows that the model with, say, $\theta=(4N+2L)^{-\Lh}$ is free from degeneracy.

As we noted after Assumption~\ref{a:nondeg}, it is expected that the energy eigenvalues of a quantum many-body system are generically nondegenerate when there is no reason (like symmetry) to cause degeneracy.
Even for a model of free fermions, we expect that possible degeneracy can be lifted by tuning some parameters, like the flux $\theta$ or the site-dependent potential or hopping amplitude.
However, it turns out that demonstrating nondegeneracy rigorously is very difficult in general.
That is why we considered an artificial setting where the system size $L$ is a prime number.
In this case, the absence of degeneracy can be demonstrated by using number-theoretic results, as we shall see below.

We also note that the absence of degeneracy was rigorously established in a disordered free fermion chain.
See Appendix~A of \cite{AbdulRahmanStolz}.

\medskip
To prove Theorems~\ref{T:nondeg} and \ref{T:nondeg2}, it is convenient to introduce the standard occupation number description.
For a given $N$-tuple $\bsk=(k_1,\ldots,k_N)$, we define the corresponding occupation numbers $\bsn=(n_{-(L-1)/2},\ldots,n_{(L-1)/2})$ as
\eq
n_\nu=\begin{cases}
1,&\text{if $2\pi\nu/L=k_j$ for some $j$};\\
0,&\text{otherwise},
\end{cases}
\en
where $\nu=0,\pm1,\ldots,\pm\Lh$.
One clearly has
\eq
\sumnu n_\nu=N.
\lb{Nn}
\en
By using the occupation numbers, the energy eigenstate \rlb{Psik} and the energy eigenvalue \rlb{Ek} are written as
\eq
\ket{\Psi_{\bsn}}=\Biggl(\prod_{\nu=-\Lh}^{\Lh}(\had_{2\pi\nu/L})^{n_\nu}\Biggr)\vac,
\en
and
\eq
E_{\bsn}=2\sumnu n_\nu\,\cos\Bigl(\frac{2\pi}{L}\nu+\theta\Bigr),
\lb{En}
\en
respectively.
By defining ``complex energy'' by
\eq
\calE_{\bsn}=\sumnu n_\nu\,\zeta^\nu,
\lb{cEn}
\en
with
\eq
\zeta=e^{i 2\pi/L},
\lb{zeta}
\en
we can express the energy eigenvalue \rlb{En} as
\eq
E_{\bsn}=2\Re[e^{i\theta}\calE_{\bsn}].
\lb{EreE}
\en

Let us state two number theoretic lemmas\footnote{
See \cite{Hal_NT} for elementary proofs of the two lemmas.
}, which play essential roles in the proof of Theorems~\ref{T:nondeg} and \ref{T:nondeg2}.
We recall $L$ is an odd prime, and $\zeta$ is defined as \rlb{zeta}.
\begin{lemma}\label{l:Gauss}
For any $m_1,\ldots,m_{L-1}\in\bbZ$ such that $m_\mu\ne0$ for some $\mu$, one has
\eq
\summu m_\mu\,\zeta^\mu\ne0.
\lb{summmune0}
\en
\end{lemma}
Here, it is crucial that the sum is from 1 to $L-1$, rather than from 1 to $L$.
Otherwise \rlb{summmune0} can never be true because $\sum_{\mu=1}^L\zeta^\mu=0$.
The lemma is a straightforward consequence of the classical result by Gauss, known as the irreducibility of the cyclotomic polynomials of prime index.
See, e.g., Chapter 12, Section 2 of \cite {Tignol}, and also Chapter 13, Section 2 of \cite{IR} or section~3.2 of \cite{StewarTall}.

The following lemma\footnote{
We learned the lemma and its proof from Wataru Kai and Kazuaki Miyatani.
} provides an explicit lower bound for $|\summu m_\mu\,\zeta^\mu|$.
\begin{lemma}\label{l:Gauss2}
For any $m_1,\ldots,m_{L-1}\in\bbZ$ such that $m_\mu\ne0$ for some $\mu$, one has
\eq
\biggl|\summu m_\mu\,\zeta^\mu\biggr|\ge\biggl(\summu|m_\mu|\biggr)^{-(L-3)/2}.
\en
\end{lemma}
\noindent
{\em Proof}\/:
The lemma is proved by using standard facts about the field norm and algebraic integers.
See, e.g., \cite{StewarTall}.
Let $\alpha=\summu m_\mu\,\zeta^\mu\in\bbZ[\zeta]\subset\bbQ[\zeta]$ and 
\eq
\sigma_j(\alpha)=\summu m_\mu\,e^{i2\pi j\mu/L},
\en
be its conjugates, where $j=1,\ldots,L-1$.
Note that $\sigma_1(\alpha)=\alpha$, $\sigma_j(\alpha)=\{\sigma_{L-j}(\alpha)\}^*$, and $|\sigma_j(\alpha)|\le M$ with $M=\summu|m_\mu|$.
Let $N:\bbQ[\zeta]\to\bbQ$ denote the field norm of $\bbQ[\zeta]$. 
By definition, we have
\eq
N(\alpha)=\prod_{j=1}^{L-1}\sigma_j(\alpha)=\prod_{j=1}^{\Lh}|\sigma_j(\alpha)|^2.
\lb{Nalpha}
\en
Since Lemma~\ref{l:Gauss} guarantees $\sigma_j(\alpha)\ne0$ for all $j$, we see that $N(\alpha)>0$.
Note that $\alpha$ is an algebraic integer, and hence so are its conjugates $\sigma_j(\alpha)$ and the norm $N(\alpha)$.
It is known that an algebraic integer that is rational must be an integer.
Since $N(\alpha)\in\bbQ$, we see $N(\alpha)\in\bbZ$ and hence $N(\alpha)\ge1$.
This bound, with \rlb{Nalpha}, implies
\eq
|\alpha|^2\ge\biggl(\prod_{j=2}^{\Lh}\bigl|\sigma_j(\alpha)\bigr|^2\biggr)^{-1}\ge\frac{1}{M^{L-3}}.\quad\qedm
\en

\medskip
We are now ready to prove our physics theorems.

\medskip
\noindent
{\em Proof of Theorem~\ref{T:nondeg}}\/:
We first show $\calE_{\bsn}\ne\calE_{\bsn'}$ if $\bsn\ne\bsn'$, where both $\bsn$ and $\bsn'$ are occupation numbers for $N$ particle energy eigenstates.
In other words, the complex energy eigenvalues are nondegenerate.
From \rlb{cEn}, we find
\eq
\calE_{\bsn}-\calE_{\bsn'}=\sumnu(n_\nu-n'_\nu)\,\zeta^\nu.
\lb{EE1}
\en
We claim that there is at least one $\nu$ such that $n_{\nu}-n'_{\nu}=0$.
To see this, it suffices to note that the converse, i.e., $n_\nu=0$, $n'_\nu=1$ or $n_\nu=1$, $n'_\nu=0$ for every $\nu$, implies $L=2N$, while $L$ is odd.
Let $\nu_0$ be such $\nu$, i.e., $n_{\nu_0}-n'_{\nu_0}=0$.
Noting that the right-hand side of \rlb{EE1} does not contain the term proportional to $\zeta^{\nu_0}$, we rewrite it as
\eq
\calE_{\bsn}-\calE_{\bsn'}=\zeta^{\nu_0}\summu m_\mu\,\zeta^\mu,
\lb{EE2}
\en
with $m_\mu=n_{\nu_0+\mu}-n'_{\nu_0+\mu}$, where we used the ``periodic boundary condition'', $\nu=\nu+L$, for the index.
Since $m_\mu$ is not identically zero (because $\bsn\ne\bsn'$), we see $\calE_{\bsn}-\calE_{\bsn'}\ne0$ from Lemma~\ref{l:Gauss}.

Now, the statement of the lemma is proved easily.
Take any $\bsn$ and $\bsn'$ with $\bsn\ne\bsn'$.
Since $e^{i\theta}(\calE_{\bsn}-\calE_{\bsn'})\ne0$, \rlb{EreE} implies that the two energy eigenvalues $E_{\bsn}$ and $E_{\bsn'}$ are degenerate only at two values of $\theta$ for which the real part of $e^{i\theta}(\calE_{\bsn}-\calE_{\bsn'})$ vanishes.
This means that the $N$-particle energy eigenvalues exhibit degeneracy at most at $\Dtot(\Dtot-1)$ different values of $\theta$, where we recalled that there are $\Dtot$ distinct $\bsn$'s.
Except for these finite points in the continuous interval $[0,2\pi)$, the Hamiltonian has no degeneracy.~\qedm

\medskip\noindent
{\em Proof of Theorem~\ref{T:nondeg2}}\/:
Consider the model with $\theta=0$.
Because of the reflection symmetry $\cos((2\pi/L)\nu)=\cos(-(2\pi/L)\nu)$, we see from \rlb{En} that the energy eigenvalue $E_{\bsn}$ depends only on $n_0$ and $n_\nu+n_{-\nu}$ for $\nu=1,\ldots,\Lh$.
In particular, we get the same energy for $n_\nu=1$, $n_{-\nu}=0$ and $n_\nu=0$, $n_{-\nu}=1$.
This means that $E_{\bsn}$ is generally degenerate, and the maximum possible degree of degeneracy is $2^N$.
We call such a degeneracy a trivial degeneracy.

We shall show that, in the model with $\theta=0$, there are no additional degeneracies than trivial degeneracies.\footnote{
\label{fn:nondegopen}
As is suggested by this conclusion, one can prove, by using essentially the same argument, the absence of degeneracy in certain open fermion chains with a suitable boundary potential.
}
Take occupation numbers $\bsn$ and $\bsn'$ for $N$ particles such that $n_\nu+n_{-\nu}\ne n'_\nu+n'_{-\nu}$ for some $\nu$ (including $\nu=0$).
The energy eigenvalues $E_{\bsn}$ and $E_{\bsn'}$ do not exhibit trivial degeneracy.
Since $\zeta^*=\zeta^{-1}$, we see from \rlb{cEn} and \rlb{EreE} that
\eq
E_{\bsn}-E_{\bsn'}=\calE_{\bsn}+(\calE_{\bsn})^*-\calE_{\bsn'}-(\calE_{\bsn'})^*
=\sumnu\tn_\nu\,\zeta^\nu,
\lb{EEEE}
\en
where we set $\tn_\nu=n_\nu+n_{-\nu}-n'_\nu-n'_{-\nu}$.
Noting that $\sumnu\zeta^\nu=0$, we rewrite \rlb{EEEE} as
\eq
E_{\bsn}-E_{\bsn'}=
\sumnu(\tn_\nu-\tn_0)\,\zeta^\nu
=\summu m_\mu\,\zeta^\mu,
\lb{EEEE2}
\en
where
\eq
m_\mu=\begin{cases}
\tn_\mu-\tn_0,&\mu=1,\ldots,\frac{L-1}{2};\\
\tn_{\mu-L}-\tn_0,&\mu=\frac{L+1}{2},\ldots,L-1.
\end{cases}
\lb{mmu}
\en
We shall see at the end of the proof that $m_\mu\ne0$ for some $\mu$.
Then, noting that 
\eq
\summu|m_\mu|\le\sumnu\{|\tn_\nu|+|\tn_0|\}\le4N+2L,
\en
we find from Lemma~\ref{l:Gauss2} that
\eq
|E_{\bsn}-E_{\bsn'}|\ge\frac{1}{(4N+2L)^{(L-3)/2}}.
\lb{EEbound}
\en
This, in particular, means that the energy eigenvalues $E_{\bsn}$ and $E_{\bsn'}$ are not degenerate.

We shall now examine the effect of nonzero $\theta$.
We make the $\theta$-dependence of the energy eigenvalues explicit by writing $E^{(\theta)}_{\bsn}$ instead of $E_{\bsn}$.

Suppose for some $\bsn\ne\bsn'$ that $E^{(0)}_{\bsn}=E^{(0)}_{\bsn'}$, i.e., $\Re\calE_{\bsn}=\Re\calE_{\bsn'}$.
The two energy eigenvalues exhibit trivial degeneracy.
Since $\calE_{\bsn}\ne\calE_{\bsn'}$ (see the proof of Theorem~\ref{T:nondeg} above), we must have $\Im\calE_{\bsn}\ne\Im\calE_{\bsn'}$.
Recalling that \rlb{EreE} implies $E^{(\theta)}_{\bsn}=2\cos\theta\,\Re\calE_{\bsn}-2\sin\theta\,\Im\calE_{\bsn}$, we see $E^{(\theta)}_{\bsn}\neq E^{(\theta)}_{\bsn'}$ for any $\theta\ne0,\pi$.
Trivial degeneracies are completely lifted.

Since we have shown that the model is free from trivial degeneracies for $\theta$ with $0<|\theta|<\pi$, we look for a sufficient condition that additional (nontrivial) degeneracy is not generated when $\theta$ is varied slightly from 0.
We observe from \rlb{En} that the resulting change in the energy eigenvalue is bounded as
\eqa
|E^{(\theta)}_{\bsn}-E^{(0)}_{\bsn}|&\le2\sumnu n_\nu\,\biggl|\cos\Bigl(\frac{2\pi}{L}\nu+\theta\Bigr)-\cos\Bigl(\frac{2\pi}{L}\nu\Bigr)\biggr|
\nl
&<2\sumnu n_\nu \,|\theta|=2N\,|\theta|,
\ena
for any $\bsn$ such that \rlb{Nn} holds.
We then find from \rlb{EEbound} that no additional degeneracy can be generated if
\eq
2\times 2N\,|\theta|\le\frac{1}{(4N+2L)^{(L-3)/2}},
\en
This is satisfied if the condition \rlb{thetacond} in the theorem is valid.

It remains to prove that $m_\mu\ne0$ for some $\mu$, where $m_\mu$ is defined in \rlb{mmu}.
To this end, we assume $m_\mu=0$ for all $\mu$.
First, suppose $\tn_0=0$.
We then have $\tn_\nu=0$ for all $\nu$, but this contradicts the basic assumption that 
$n_\nu+n_{-\nu}\ne n'_\nu+n'_{-\nu}$ for some $\nu$.
Next, suppose $\tn_0\ne0$.
We then have $n_\nu+n_{-\nu}-n'_\nu-n'_{-\nu}=\tn_0\ne0$ for any $\nu\ne0$.
But this implies $\sum_{\nu\ne0}n_\nu-\sum_{\nu\ne0}n'_\nu=\frac{L-1}{2}\tn_0$, which apparently contradicts with the constraint on the total particle number, i.e., $\sum_{\nu}n_\nu=\sum_{\nu}n'_\nu=N$.~\qedm

\subsection{Justification of Assumption~\ref{a:P2N}}\label{s:P2N}
We shall demonstrate that Assumption~\ref{a:P2N} about the particle distribution in the energy eigenstates is valid in the present free fermion chain.
As in section~\ref{s:setting}, we disjointly decompose the chain $\La=\{1,\ldots,L\}$ as $\La=\La_1\cup\La_2$ with $|\La_1|=(L-1)/2$ and $|\La_2|=(L+1)/2$.
An obvious choice is $\La_1=\{1,2,\ldots,(L-1)/2\}$, but any subset will work similarly.

Let us decompose the creation operator $\had_k$ defined in \rlb{ak} as 
\eq
\had_k=\hbd_{1,k}+\hbd_{2,k}, 
\lb{aaa}
\en
where
\eq
\hbd_{\alpha,k}=\frac{1}{\sqrt{L}}\sum_{x\in\La_\alpha}e^{ikx}\,\hcd_x,
\en
with $\alpha=1,2$.
Note that $\{\hbd_{1,k},\hb_{1,k'}\}$ with $k\ne k'$ is not necessarily vanishing.
From \rlb{Psik}, we obviously have
\eq
\hP_1\ket{\Psi_{\bsk}}=\hbd_{1,k_1}\hbd_{1,k_2}\ldots\hbd_{1,k_N}\vac,
\en
and hence
\eqa
\bra{\Psi_{\bsk}}\hP_1\ket{\Psi_{\bsk}}&=\vacb\hb_{1,k_N}\ldots\hb_{1,k_2}\hb_{1,k_1}
\hbd_{1,k_1}\hbd_{1,k_2}\ldots\hbd_{1,k_N}\vac
\nl&\le \snorm{\hb_{1,k_1}\hbd_{1,k_1}}\,\vacb\hb_{1,k_N}\ldots\hb_{1,k_2}
\hbd_{1,k_2}\ldots\hbd_{1,k_N}\vac.
\lb{PPP}
\ena
Here we used the basic property $\bra{\Psi}\hat{A}\ket{\Psi}\le \snorm{\hat{A}}\sbkt{\Psi|\Psi}$ of the operator norm of an arbitrary operator $\hat{A}$.
Noting that
\eq
\snorm{\hb_{1,k}\hbd_{1,k}}\le\frac{1}{2},
\lb{aa12}
\en
for any $k\in\calK$ (as we shall show below), we get the desired bound \rlb{2N} by repeatedly using \rlb{PPP}.

To estimate the norm $\snorm{\hb_{1,k}\hbd_{1,k}}$, we first note by an explicit calculation that $\{\hb_{1,k}, \hbd_{1,k}\}=p$ with $p=(L-1)/(2L)\le 1/2$.
Then by noting that $(\hb_{1,k}\hbd_{1,k})^2=p\,\hb_{1,k}\hbd_{1,k}$, we see that the self-adjoing operator $\hb_{1,k}\hbd_{1,k}$ has eigenvalues 0 and $p$.
This means $\snorm{\hb_{1,k}\hbd_{1,k}}=p$, which implies \rlb{aa12}.

It is clear that the above justification of Assumption~\ref{a:P2N} applies to a much more general class of free fermion systems.
The only requirement is that there is a decomposition corresponding to \rlb{aaa} of the creation operator for single-particle energy eigenstate with the property \rlb{aa12}.
See Appendix \ref{s:A2} for a class of examples.

\section{Discussion}
\label{s:discussion}
We developed in section~\ref{s:main} a general theory for the thermalization in low-density lattice gases.
Under the two essential assumptions, namely, Assumption~\ref{a:P2N} about the particle distribution in energy eigenstates and Assumption~\ref{a:nondeg} about nondegneracy of energy eigenvalues, we have shown that the system exhibits thermalization (in a restricted sense) when the initial state is drawn randomly from the Hilbert space $\calH_1$ in which all particles are confined in the half-lattice $\La_1$.
The essential observation, which is summarized in Theorem~\ref{t:Deff}, is that Assumptions~\ref{a:P2N} implies the lower bound \rlb{DDbound} on the effective dimension of the initial state.
Combined with standard arguments, the lower bound implies the desired statement about thermalization.

Then, in section~\ref{s:freefermion}, we justified Assumptions~\ref{a:nondeg} and \ref{a:P2N} for a class of free fermion chains without relying on any unproven assumptions.
Free fermion models, which have infinitely many conserved quantities, are often referred to as examples of systems that fail to thermalize.
One might then be puzzled to see that we have established thermalization in free fermion chains.
The essential point is in the choice of the initial state $\kPz$.
In a non-interacting fermion model with translation invariance, for example, the momentum distribution does not change under the unitary time evolution.
Thus the system never thermalizes if it starts from a state with non-thermal momentum distribution.
In our case, the momentum distribution is thermal from the beginning because the initial state is chosen to be a thermal state but with particles confined in the sublattice $\La_1$.
In the language of generalized Gibbs ensembles, our generalized ensemble with the random initial state is characterized by local integrals of motion taking the same values as the equilibrium ones.

\medskip

Although free fermion chains are the only examples in which we can fully justify our two assumptions, we stress that our general theory should apply to much more general models, most of which are non-integrable.
Non-integrable models are believed to exhibit robust thermalization from an arbitrary realistic nonequilibrium initial state.
When applied to such models, our thermalization theorem is expected to describe a partial aspect of thermalization exhibited by the model.
We might say that our theory is general and broad enough to cover not only full-fledged thermalization in non-integrable system but also (rather trivial) thermalization in free fermion chains.

As we have already discussed after Assumption~\ref{a:nondeg}, it is believed that energy eigenvalues are nondegenerate in a generic non-integrable model.
Therefore, let us focus on Assumption~\ref{a:P2N}, which asserts that the probability of finding all particles in the sublattice $\La_1$ does not exceed $2^{-N}$ in any energy eigenstate as in \rlb{2N}.
By accepting the assumption of nondegeneracy as a plausible one, we have two additonal classes of models in which we can prove \rlb{2N} as presented in Appendix~\ref{s:doublelattice}.

Assumption~\ref{a:P2N} is reminiscent of the (strong) ETH in the sense that we postulate that every energy eigenstate exhibits more or less uniform particle distributions.
Although we are able to prove the bound \rlb{2N} only for limited models, we expect that it is valid for all (or for a great majority of) energy eigenstates of a generic macroscopic quantum system.
The bound does not hold, for example, in a state where a macroscopic number of particles form a big cluster and move together, but such states cannot be an energy eigenstate of a model with short-range interactions.
We, in particular, note that the average of the probability $\bPj\hP_1\kPj$ over all the energy eigenstates is
\eq
\frac{1}{\Dtot}\sum_{j=1}^{\Dtot}\bPj\hP_1\kPj
=\frac{1}{\Dtot}\Tr[\hP_1]=\frac{D_1}{\Dtot}\sim 2^{-N}e^{-(L/2)\rho^2},
\en
and is much smaller than $2^{-N}$.

\medskip

In section~\ref{s:main}, we only discussed thermalization in the sense that the ratio of the number of particles in a macroscopic region $\Gamma$ approaches its equilibrium value $\gamma$.
It is, however, clear from the proof that our method automatically extends to other criteria for thermal equilibrium.
Let $\hP_\mathrm{neq}$ be the projection operator onto the nonequilibrium subspace of $\Htot$ determined by a certain criterion for thermal equilibrium.
If the canonical expectation value of $\hP_\mathrm{neq}$ at infinite temperature satisfies
\eq
\sbkt{\hP_\mathrm{neq}}_\infty\le e^{-\kappa N}=e^{-\kappa\rho\,L},
\lb{LD}
\en
with a constant $\kappa$ such that $\kappa>\rho$, then we can prove, exactly as in Theorem~\ref{t:main}, that the expectation value $\bPt\hP_\mathrm{neq}\kPt$ is extremely small for sufficiently large typical $t$, i.e., the system exhibits thermalization.
Although we do not go into details, we expect that the assumption \rlb{LD} is valid if one defines $\hP_\mathrm{neq}$ to be the projection onto the space where the total energy in a macroscopic region differs considerably from its expectation value in $\sbkt{\cdot}_\infty$.

We note, however, that if one employs a criterion of thermal equilibrium that involves, say, particle-particle correlation, then the assumption \rlb{LD} with $\kappa>\rho$ for the corresponding nonequilibrium projection is never valid.
This means that our theorem is simply powerless.
This shortcoming is related to the limitation to low densities and reflects the limitation of our approach, which reflects our strategy to base the theory on mild assumptions.

\appendix

\bigskip\bigskip\bigskip
\noindent
{\bf \LARGE Appendices}

\section{Models with degenerate energy eigenvalues}
\label{s:deg}
Our general discussion in section~\ref{s:main} is based on the crucial assumption, Assumption~\ref{a:nondeg}, that all the energy eigenvalues are nondegenerate.
Here we shall see how one can treat models in which the degree of degeneracy is at most $d_\mathrm{max}$.
We find that our thermalization results remain valid as long as $d_\mathrm{max}$ is not too large.
Unfortunately, we do not know of any examples where a nontrivial upper bound for the degree of degeneracy is known.

Let $E_j$ with $j=1,\ldots,\Nel$ be the distinct energy eigenvalues.
We denote by $\ket{\Psi_{j,\ell}}$ with $\ell=1,\ldots,d_j$ the energy eigenstates corresponding to $E_j$, where $d_j$ is the degree of degeneracy of $E_j$.
We assume that the collection of $\ket{\Psi_{j,\ell}}$ with all $j$, $\ell$ forms an orthonormal basis of $\Htot$.

We first examine the discussion in section~\ref{s:Deff} about the effective dimension.
A straightforward generalization of the definition \rlb{Deff} of the effective dimension is
\eq
\Deff=\biggl(\sum_{j=1}^{\Nel}\sum_{\ell=1}^{d_j}\bigl|\bkt{\Phi(0)|\Psi_{j,\ell}}\bigr|^4\biggr)^{-1}.
\lb{Deffdeg1}
\en
When energy eigenvalues are degenerate, however, it is convenient to employ the definition
\eq
\tDeff=\biggl(\sum_{j=1}^{\Nel}\bra{\Phi(0)}\hP_j\ket{\Phi(0)}^2\biggr)^{-1},
\lb{Deffdeg2}
\en
where $\hP_j=\sum_{\ell=1}^{d_j}\ket{\Psi_{j,\ell}}\bra{\Psi_{j,\ell}}$ is the projection onto the space corresponding to the energy eigenvalue $E_j$.
Clearly, \rlb{Deffdeg2} reduces to the original \rlb{Deffdeg1} when there is no degeneracy.
To evaluate \rlb{Deffdeg2}, we note that
\eqa
\bra{\Phi(0)}\hP_j\ket{\Phi(0)}^2&=\sum_{\ell,\ell'=1}^{d_j}\bigl|\bkt{\Phi(0)|\Psi_{j,\ell}}\bigr|^2\,\bigl|\bkt{\Phi(0)|\Psi_{j,\ell'}}\bigr|^2
\nl&\le\frac{1}{2}\sum_{\ell,\ell'=1}^{d_j}\Bigl(\bigl|\bkt{\Phi(0)|\Psi_{j,\ell}}\bigr|^4+\bigl|\bkt{\Phi(0)|\Psi_{j,\ell'}}\bigr|^4\Bigr)
\nl&=d_j\sum_{\ell=1}^{d_j}\bigl|\bkt{\Phi(0)|\Psi_{j,\ell}}\bigr|^4,
\ena
where we noted $ab\le(a^2+b^2)/2$ to get the second line.
We thus find
\eq
\tDeff^{-1}\le\sum_{j=1}^{\Nel}d_j\sum_{\ell=1}^{d_j}\bigl|\bkt{\Phi(0)|\Psi_{j,\ell}}\bigr|^4
\le d_\mathrm{max}\sum_{j=1}^{\Nel}\sum_{\ell=1}^{d_j}\bigl|\bkt{\Phi(0)|\Psi_{j,\ell}}\bigr|^4
=\frac{d_\mathrm{max}}{\Deff}
\lb{Deffdeg3}
\en
where $d_\mathrm{max}=\max_jd_j$.

Suppose that \rlb{2N} in Assumption~\ref{a:P2N} is valid for the energy eigenstates $\ket{\Psi_{j,\ell}}$.
Then Theorem~\ref{t:Deff} guarantees the crucial lower bound \rlb{DDbound} for $\Deff$ defined as \rlb{Deffdeg1}.
We thus find from \rlb{Deffdeg3} that
\eq
\frac{\Dtot}{\tDeff}\le d_\mathrm{max}\,e^{\rho N}.
\en
We see that $\tDeff$ is large provided that $d_\mathrm{max}$ is not too large.
Note that the degeneracy does not essentially change the behavior of the effective dimension if  $d_\mathrm{max}$ grows subexponentially in $N$.

We move onto the discussion in section~\ref{s:TET} about the time evolution.
Taking into account the degeneracy, the expression \rlb{Phit} for the time evolution reads
\eq
\kPt=e^{-i\hH t}\kPz
=\sum_{j=1}^{\Nel}e^{-iE_jt}\hP_j\ket{\Phi(0)}
=\sum_{j=1}^{\Nel}e^{-iE_jt}\,\ket{\widetilde{\Psi}_j}\,
\sqrt{\bra{\Phi(0)}\hP_j\ket{\Phi(0)}},
\lb{Phit2}
\en
where we defined
\eq
\ket{\widetilde{\Psi}_j}=\frac{\hP_j\ket{\Phi(0)}}{\snorm{\hP_j\ket{\Phi(0)}}}.
\en
Correspondingly, \rlb{PPP1} is modified as
\eqa
\lim_{T\up\infty}\frac{1}{T}\int_0^Tdt\,\bPt\Pneq\kPt
&=\sum_{j=1}^{\Nel}\bra{\Phi(0)}\hP_j\ket{\Phi(0)}\bra{\widetilde{\Psi}_j}\Pneq\ket{\widetilde{\Psi}_j}
\nl&\le\sqrt{
\biggl(\sum_{j=1}^{\Nel}\bra{\Phi(0)}\hP_j\ket{\Phi(0)}^2\biggr)
\biggl(\sum_{j=1}^{\Nel}\bra{\widetilde{\Psi}_j}\Pneq\ket{\widetilde{\Psi}_j}^2\biggr)
}
\nl&\le\sqrt{
\biggl(\sum_{j=1}^{\Nel}\bra{\Phi(0)}\hP_j\ket{\Phi(0)}^2\biggr)
\biggl(\sum_{j=1}^{\Nel}\sum_{\ell=1}^{d_j}\bra{\Psi_{j,\ell}}\Pneq\ket{\Psi_{j,\ell}}^2\biggr)
}
\nl&\le\sqrt{
\Dtot\tDeff^{-1}\,\sbkt{\Pneq}_\infty.
}
\lb{PPP1deg}
\ena
Therefore the rest of the discussion remains valid if we replace $\Deff$ with $\tDeff$.

\section{Models satisfying Assumption~\ref{a:P2N}}
\label{s:doublelattice}
In this Appendix, we present two classes of models in which we can prove Assumption~\ref{a:P2N} about the particle distribution (under suitable assumptions about nondegeneracy).
If we could also justify Assumption~\ref{a:nondeg} about the nondegeneracy of energy eigenvalues, we would have further rigorous examples of thermalization.
Unfortunately, we still do not know how nondegeneracy can be proved, although we believe it to be highly plausible.

The first class of models is that of interacting fermions on a specific class of lattices, while the second class is that of free fermions on arbitrary lattices with  $\bbZ_2$ symmetry.

\subsection{Interacting fermions on double-lattice}
\label{s:A1}
First, we discuss a class of lattice gas models on a double lattice with special symmetry.
In these models, we can easily verify the bound \rlb{2N} for any energy eigenstate corresponding to a nondegenerate energy eigenvalue.
See Lemma~\ref{L:DL} below.
This means that Assumption~\ref{a:P2N} about the particle distribution is automatically valid if Assumption~\ref{a:nondeg} about nondegeneracy of energy eigenvalues is valid.
The class of models in fact contains many non-trivial interacting models for which we generally expect that energy eigenvalues are nondegenerate.
We thus expect that the present class contains many examples in which our thermalization theorem, Theorem~\ref{t:main}, is valid.
Unfortunately, we are not able to prove nondegeneracy in concrete models, except for trivial decoupled models.
See the discussion at the end of the present subsection.

We shall describe the class of models and state the basic observation, i.e., Lemma~\ref{L:DL}.
Although we here describe models of fermions for notational simplicity, extensions to hardcore bosons or quantum spin systems are trivial.

Let $\La_0$ be a lattice with $L/2$ sites, and $\La_1$ and $\La_2$ be copies of $\La_0$.
Sites in $\La_1$ and $\La_2$ are denoted as $(x,1)$ and $(x,2)$, respectively, with $x\in\La_0$.
We consider a model of fermions on the whole lattice $\La=\La_1\cup\La_2$.

We assume that the Hamiltonian $\hH$ conserves the total particle number and is invariant under the exchange of two sites $(x,1)$ and $(x,2)$ for each $x\in\La_0$.
The latter is a highly nontrivial (and artificial) assumption, which enables us to prove the desired bound \rlb{2N} easily.
To be more precise we define for each $x\in\La_0$ the unitary operator $\hU_x$ that swaps $(x,1)$ and $(x,2)$.
It is defined by $\hU_x\vac=\vac$, $\hU_x\,\hc_{(x,1)}\hU_x^\dagger=\hc_{(x,2)}$, $\hU_x\,\hc_{(x,2)}\hU_x^\dagger=\hc_{(x,1)}$, and $\hU_x\,\hc_{(y,\nu)}\hU_x^\dagger=\hc_{(y,\nu)}$ for $y\ne x$ and $\nu=1,2$.
Note that $(\hU_x)^2=\hat{1}$.
Our symmetry assumption is that $[\hU_x,\hH]=0$ for any $x\in\La_0$.

If we restrict ourselves to models with standard particle-hopping and two-body interactions, the most general Hamiltonian takes the form
\eqa
\hH&=\sumtwo{x,y\in\La_0}{(x\ne y)}\bigl\{t_{x,y}(\hcd_{(x,1)}+\hcd_{(x,2)})(\hc_{(y,1)}+\hc_{(y,2)})
+\frac{v_{x,y}}{2}(\hn_{(x,1)}+\hn_{(x,2)})(\hn_{(y,1)}+\hn_{(y,2)})\bigr\}
\nl&+\sum_{x\in\La_0}\bigl\{s_x(\hcd_{(x,1)}\hc_{(x,2)}+\hcd_{(x,2)}\hc_{(x,1)})+w_x(\hn_{(x,1)}+\hn_{(x,2)})+u_x\,\hn_{(x,1)}\hn_{(x,2)}\bigr\},
\lb{Hdouble}
\ena
where $t_{x,y}=(t_{y,x})^*\in\bbC$, $v_{x,y}=v_{y,x}\in\bbR$, and $s_x,w_x,u_x\in\bbR$.
We defined the number operator by $\hn_{(x,\sigma)}=\hcd_{(x,\sigma)}\hc_{(x,\sigma)}$.

\medskip
Here is the basic observation in the present appendix.
\begin{lemma}\label{L:DL}
Let $\ket{\Psi}$ be the normalized eigenstate corresponding to a nondgenerate energy eigenvalue of $\hH$.
Then we have
\eq
\bra{\Psi}\hP_1\ket{\Psi}\le2^{-N},
\lb{2NPsi}
\en
which is the same as \rlb{2N}.
\end{lemma}

\noindent{\em Proof:}\/
For a fixed particle number $N$, we define the basis states of the model by
\eq
\ket{\Xi_{S_1,S_2}}=\Bigl(\prod_{x\in S_1}\hcd_{(x,1)}\Bigr)\Bigl(\prod_{x\in S_2}\hcd_{(x,2)}\Bigr)\vac,
\en
where $S_1$ and $S_2$ are arbitrary subsets of $\La_0$ such that $|S_1|+|S_2|=N$.
Take any normalized eigenstate $\ket{\Psi}$ of $\hH$ and expand it in the above basis as
\eq
\ket{\Psi}=\sumtwo{S_1,S_2\subset\La_0}{(|S_1|+|S_2|=N)}\psi_{S_1,S_2}\ket{\Xi_{S_1,S_2}},
\en
where $\psi_{S_1,S_2}\in\bbC$ are coefficients which satisfy $\sum|\psi_{S_1,S_2}|^2=1$.
The symmetry of the Hamiltonian and the nondgeneracy imply $\hU_x\ket{\Psi}=\pm\ket{\Psi}$ for any $x\in\La_0$.
This means that the expansion coefficients satisfy
\eq
|\psi_{S,\emptyset}|=|\psi_{S\backslash S',S'}|,
\en
for any $S\subset\La_0$ with $|S|=N$ and any $S'\subset S$.
We thus have
\eq
|\psi_{S,\emptyset}|^2=\frac{1}{2^N}\sum_{S'\subset S}|\psi_{S\backslash S',S'}|^2,
\en
which, when summed over $S$, yields
\eq
\sumtwo{S\subset\La_0}{(|S|=N)}|\psi_{S,\emptyset}|^2=\frac{1}{2^N}\sumtwo{S\subset\La_0}{(|S|=N)}\sum_{S'\subset S}|\psi_{S\backslash S',S'}|^2\le\frac{1}{2^N}\sumtwo{S_1,S_2\subset\La_0}{(|S_1|+|S_2|=N)}|\psi_{S_1,S_2}|^2=\frac{1}{2^N}.
\en
Noting that the left-hand side is $\bra{\Psi}\hP_1\ket{\Psi}$, we get \rlb{2NPsi}.~\qedm

\medskip
This model considered here is generally non-integrable, and we expect that its energy eigenvalues are nondegenerate.
It is desirable to find models in which the absence of degeneracy can be proved rigorously.

Unfortunately, the only case we can prove nondegeneracy is a trivial decoupled model with $t_{x,y}=v_{x,y}=0$ for all $x,y\in\La_0$.
We readily see that the energy eigenvalues are nondegenerate if $s_x$, $w_x$, and $u_x$ with $x\in\La_0$ are chosen to be different from each other.\footnote{
\label{fn:decoupled}
In this trivial model, the energy eigenvalues for a pair of sites $(x,1)$ and $(x,2)$ are either zero (when there is no particles), $\pm s_x+w_x$ (when there is one particle), or $2w_x+u_x$ (when there are two particles).
The total energy eigenvalues are the sums of these eigenvalues and are nondegenerate if we choose $s_x$, $w_x$, and $u_x$ properly.
}
Therefore we can fully justify our main theorem, Theorem~\ref{t:main}, for the model, but we should note that the result is trivial.
In the initial state $\kPz$, each pair of sites $(x,1)$ and $(x,2)$ is either empty or occupied by one partilce at $(x,1)$.
The time evolution then takes place independently for each pair of sites.
If there is a particle in a pair, then a superposition of two states with a particle at $(x,1)$ and at $(x,2)$ is generated.
This, when viewed macroscopically, results in thermalization.
We must say that there is nothing interesting in this observation.

\subsection{Free fermions with $\bbZ_2$ symmetry}
\label{s:A2}
Next, we discuss a class of free fermion models in which the bound \rlb{2N} for the particle distribution, and hence Assumption~~\ref{a:P2N} can be justified.
We here follow the strategy outlined at the end of section~\ref{s:P2N} and justify the inequality \rlb{aa12} for the fermion operators corresponding to single-particle energy eigenstates.

Let $\La$ be an arbitrary lattice, and consider the most general free fermion Hamiltonian
\eq
\hH=\sum_{x,y\in\La}t_{x,y}\,\hcd_x\hc_y,
\en
where the hopping amplitude satisfies $t_{x,y}=(t_{y,x})^*\in\bbC$.
Note that the diagonal element $t_{x,x}\in\bbR$ represents the single-body potential.

We assume that the model has $\bbZ_2$ symmetry in the sense that there is a one-to-one map $p:\La\to\La$ such that $p^2={\rm id}$, and that the Hamiltonian is invariant under the transformation $p$, i.e., $t_{p(x),p(y)}=t_{x,y}$ for any $x,y\in\La$.
We also assume that $\La$ is disjointly decomposed as $\La=\La_1\cup\La_2$ and that $p(\La_1)\subset\La_2$.

As an example, consider the chain $\La=\{1,\ldots,L\}$ with odd $L$, and let $p$ be the inversion $p(x)=L+1-x$.
Then the decomposition with $\La_1=\{1,\ldots,(L-1)/2\}$ and $\La_2=\{(L+1)/2,\ldots,L\}$ satisfies the above property.

Let $\bspsi=(\psi_x)_{x\in\La}$ be a normalized single-particle energy eigenstate.
To be precise, it satisfies the Schr\"odinger equation $\epsilon\,\psi_x=\sum_{y\in\La}t_{x,y}\psi_x$ for all $x\in\La$ with the (single-particle) energy eigenvalue $\epsilon$.
Let us further assume that the energy eigenvalue $\epsilon$ is nondegenerate.
Then, with respect to the symmetry transformation $p$,  the corresponding wave function $\bspsi$ is either symmetric, i.e., $\psi_{p(x)}=\psi_x$ for all $x\in\La$, or antisymmetric, i.e.,  $\psi_{p(x)}=-\psi_x$ for all $x\in\La$.
We then see that
\eq
\sum_{x\in\La_1}|\psi_x|^2=\frac{1}{2}\sum_{x\in\La_1}\bigl(|\psi_x|^2+|\psi_{p(x)}|^2\bigr)
\le\frac{1}{2}\sum_{x\in\La}|\psi_x|^2=\frac{1}{2},
\lb{La1psi}
\en
where we noted that $p(\La_1)\subset\La\backslash\La_1$.

Let $\had_{\bspsi}=\sum_{x\in\La}\psi_x\,\hcd_x$ be the creation operator of the state $\bspsi$.
It can be decomposed as $\had_{\bspsi}=\hbd_{1,\bspsi}+\hbd_{2,\bspsi}$ with $\hbd_{1,\bspsi}=\sum_{x\in\La_1}\psi_x\,\hcd_x$ and $\hbd_{2,\bspsi}=\sum_{x\in\La_2}\psi_x\,\hcd_x$.
This corresponds to the decomposition \rlb{aaa}.
We also see from \rlb{La1psi} that $\hbd_{1,\bspsi}$ satisfies $\snorm{\hb_{1,\bspsi}\,\hbd_{1,\bspsi}}=\sum_{x\in\La_1}|\psi_x|^2\le1/2$, which corresponds to the desired \rlb{aa12}.

We now assume that single-particle energy eigenvalues $\epsilon_1,\ldots,\epsilon_{|\La|}$ are all nondegenerate, and denote by $\had_j$ the creation operator of the single-particle energy eigenstate corresponding to $\epsilon_j$.
Then the foregoing discussion shows that each $\had_j$ is decomposed as \rlb{aaa}, and the operator for the sublattice $\La_1$ satisfies the bound \rlb{aa12}.
Repeating the derivation in section~\ref{s:P2N}, we see an $N$-body energy eigenstate of the form
\eq
\ket{\Psi}=\had_{j_1}\cdots\had_{j_N}\vac,
\en
satisfies the desired bound \rlb{2N}.

Interestingly, it was only necessary to assume the nondegeneracy of single-particle energy eigenvalues to prove the desired bound \rlb{2N} in this model.
To ensure the presence of thermalization, we have to assume further that $N$-body energy eigenvalues are nondegenerate.
It is rather likely that degeneracy is absent in a sufficiently complex free fermion model, but we do not know how to justify the claim.
We also note that the $p$-symmetry may not be exact.  It can be violated by a small perturbation as long as the bound \rlb{2N} remains valid.

\section{Effective dimensions of some initial particle configurations in the free fermion chain}
\label{s:Deffx}
In the main text, the initial state $\kPz$ is drawn randomly from the small Hilbert space $\calH_1$.
Conceptually speaking, it may be desirable to consider the time evolution starting from a non-random simple initial state.
Here we again treat free fermion chains and examine the effective dimensions of some initial states in which particles have definite positions.

In section~\ref{s:periodic}, we observe that the initial state where particles are arranged in a periodic manner has an effective dimension that is large but not large enough to guarantee thermalization.
This observation suggests that a random initial state is mandatory in a free fermion model if we demand the effective dimension to be extremely large.
This is very likely to be a common property for integrable models.
In a non-integrable model, on the other hand, it is believed that even a regular initial state generally has an effective dimension almost as large as the total dimension.

In section~\ref{s:Golomb}, we consider an artificial class of initial configurations (Golomb ruler configurations) and show that the corresponding effective dimensions are almost as large as the total dimension.
This leads to a statement about thermalization with a non-random initial state.
In this class of models, however, the particle density inevitably tends to zero according to $\rho\sim N^{-1}$ as the particle number grows.

\subsection{General formula for $\Deff$}
We consider the free fermion chain as defined in section~\ref{s:freefermion}.
Let the initial particle configuration be $\bsx=(x_1,x_2,\ldots,x_N)$ with $x_j\in\{1,\ldots,L\}$ such that $x_j<x_{j+1}$ for $j=1,\ldots,N-1$, and define the corresponding $N$ fermion state as
\eq
\ket{\Phi_{\bsx}}=\hcd_{x_1}\hcd_{x_2}\cdots\hcd_{x_N}\vac.
\en
We set $\ket{\Phi_{\bsx}}$ as the initial state $\kPz$.
Then we see from \rlb{Deff} that the effective dimension is given by
\eq
\Deff^{-1}=\sum_{\bsk\in\tilde{\calK}_N}\bigl|\bra{\Phi_{\bsx}}\Psi_{\bsk}\rangle\bigr|^4,
\lb{Deff3}
\en
where $\tilde{\calK}_N=\{(k_1,\ldots,k_N)\,|\,k_j<k_{j+1}\}\subset\calK^N$.
(The $k$-space $\calK$ is defined in \rlb{K}.)
Noting that \rlb{ak} implies $\{\hc_x,\had_k\}=e^{ikx}/\sqrt{L}$, we see
\eq
\bra{\Phi_{\bsx}}\Psi_{\bsk}\rangle=\vacb\hc_{x_N}\cdots\hc_{x_1}\had_{k_1}\cdots\had_{k_N}\vac
=L^{-N/2}\sum_P(-1)^P\prod_{j=1}^Ne^{ik_jx_{P(j)}},
\lb{PhiPsi}
\en
where the summation is over all possible $N!$ permutations $P$ of $\{1,\ldots,N\}$ and $(-1)^P$ is the signature of $P$.
It is useful to regard $\bsk$ in the above expression as an element in $\calK^N$ rather than its physical subspace $\tilde{\calK}_N$.
This replacement is justified since $\bigl|\bra{\Phi_{\bsx}}\Psi_{\bsk}\rangle\bigr|$ is invariant under any permutations of $k_1,\ldots,k_N$ and equals zero if $k_j=k_{j'}$ for some $j\ne j'$.
We can thus rewrite \rlb{Deff3} as
\eq
\Deff^{-1}=\frac{1}{N!}\sum_{\bsk\in\calK^N}\bigl|\bra{\Phi_{\bsx}}\Psi_{\bsk}\rangle\bigr|^4.
\lb{Dsum}
\en
This rewriting is useful since one can now sum independently over $k_1,\ldots,k_N\in\calK$.

From \rlb{PhiPsi}, we see that
\eqa
\bigl|\bra{\Phi_{\bsx}}\Psi_{\bsk}\rangle\bigr|^2&=
\frac{1}{L^N}\sum_{P,Q}(-1)^{PQ}\prod_{j=1}^Ne^{ik_j(x_{P(j)}-x_{Q(j)})}
\nl&=\frac{N!}{L^N}
+\frac{1}{L^N}\sumtwo{P,Q}{(P\ne Q)}(-1)^{PQ}\prod_{j=1}^Ne^{ik_j(x_{P(j)}-x_{Q(j)})},
\lb{PhiPsi2}
\ena
and
\eq
\bigl|\bra{\Phi_{\bsx}}\Psi_{\bsk}\rangle\bigr|^4=C_1+C_2(\bsk)+C_3(\bsk),
\lb{C1C2C3}
\en
with
\eqg
C_1=\Bigl(\frac{N!}{L^N}\Bigr)^2,\quad
C_2(\bsk)=\frac{2N!}{L^{2N}}\sumtwo{P,Q}{(P\ne Q)}(-1)^{PQ}\prod_{j=1}^Ne^{ik_j(x_{P(j)}-x_{Q(j)})},\lb{C1C2}\\
C_3(\bsk)=\frac{1}{L^{2N}}\sumtwo{P,Q,P',Q'}{(P\ne Q,\,P'\ne Q')}(-1)^{PQP'Q'}
\prod_{j=1}^Ne^{ik_j\{(x_{P(j)}-x_{Q(j)})-(x_{P'(j)}-x_{Q'(j)})\}}.
\lb{C3}
\eng
 We shall evaluate the sum \rlb{Dsum} by using the decomposition \rlb{C1C2C3}.
Clearly 
\eq
\frac{1}{N!}\sum_{\bsk\in\calK^N}C_1=\frac{N!}{L^N},
\lb{C1sum}
\en
 The remaining sums are evaluated by using the standard formula
\eq
\sum_{k\in\calK}e^{ikx}=\begin{cases}
L,&x=0\mod L;\\
0,&\text{otherwise},
\end{cases}
\en
where $x\in\bbZ$.
Note that in the expression for $C_2(\bsk)$ in \rlb{C1C2}, one has $x_{P(j)}-x_{Q(j)}\ne0$ for at least one $j$ because $P\ne Q$.
We thus see
\eq
\frac{1}{N!}\sum_{\bsk\in\calK^N}C_2(\bsk)=0.
\lb{C2sum}
\en
The sum of $C_3(\bsk)$ is generally nonzero and can be evaluated as
\eqa
\frac{1}{N!}&\sum_{\bsk\in\calK^N}C_3(\bsk)
\nl&=
\frac{1}{L^NN!}\sumtwo{P,Q,P',Q'}{(P\ne Q,\,P'\ne Q')}(-1)^{PQP'Q'}
\prod_{j=1}^N\chi[(x_{P(j)}-x_{Q(j)})-(x_{P'(j)}-x_{Q'(j)})=0\mod L],
\lb{C3sum}
\ena
where the characteristic function is defined as $\chi[\text{true}]=1$ and $\chi[\text{false}]=0$.
Let us write the right-hand side of \rlb{C3sum} as $\calS_{\bsx}/L^N$.
From \rlb{Dsum}, \rlb{C1C2C3}, \rlb{C1sum}, \rlb{C2sum}, and \rlb{C3sum}, we see that the effective dimension of the initial state $\kPz=\ket{\Phi_{\bsx}}$ is given by
\eq
\Deff=\frac{L^N}{N!+\calS_{\bsx}}.
\lb{DeffS}
\en

Our main task is to evaluate the sum $\calS_{\bsx}$ defined in \rlb{C3sum} for a given particle configuration $\bsx$.
For later convenience we sum over $P$ in \rlb{C3sum} (and write $P^{-1}Q$, $P^{-1}P'$, and $P^{-1}Q'$ as $Q$, $P'$, and $Q'$, respectively) to rewrite the expression as
\eq
\calS_{\bsx}=\sumtwo{Q,P',Q'}{(Q\ne\mathrm{id},\,P'\ne Q')}(-1)^{QP'Q'}
\prod_{j=1}^N\chi[x_{j}-x_{Q(j)}-x_{P'(j)}+x_{Q'(j)}=0\mod L].
\lb{Sx}
\en

\subsection{$\Deff$ in periodic configurations}
\label{s:periodic}
First, we consider regular particle configurations with a period $p=1,2,\ldots$.
Fix $p$, and choose the chain length $L$ and the particle number $N$ such that $L=pN$.
We consider the initial particle distribution given by
\eq
x_j=pj,
\en
for $j=1,\ldots,N$.

Then \rlb{Sx} is computed as
\eqa
\calS_{\bsx}&=\sumtwo{Q,P',Q'}{(Q\ne\mathrm{id},\,P'\ne Q')}(-1)^{QP'Q'}
\prod_{j=1}^N\chi[pj-{pQ(j)}-{pP'(j)}+{pQ'(j)}=0\mod L]
\nl&=\sumtwo{Q,P',Q'}{(Q\ne\mathrm{id},\,P'\ne Q')}(-1)^{QP'Q'}
\prod_{j=1}^N\chi[j-{Q(j)}-{P'(j)}+{Q'(j)}=0\mod N],
\lb{Sxper}
\ena
which depends only on $N$ and is independent of $L$ and $p$.
Thus, we can evaluate the above sum by employing a useful choice of $p$.
Fortunately, this sum becomes trivial for $p=1$, and therefore we compute the sum in the case of $p=1$.
A fermion system with $L=N=1$, which is fully filled, has one-dimensional Hilbert space and hence $\Deff=1$.
We see from \rlb{DeffS} that $\calS_{\bsx}=N^N-N!$.
Recalling the $L$ independence of $\calS_{\bsx}$, we get a remarkably simple result
\eq
\Deff=\Bigl(\frac{L}{N}\Bigr)^N=e^{-(\rho\log\rho)L},
\lb{Deffp}
\en
for any $L$ and $N$ (such that $L=pN$), where $\rho=1/p$ is the particle density.
We thus see that the effective dimension grows exponentially with the system size $L$, as expected.
But it turns out that it is not large enough.
Combining \rlb{Deffp} with \rlb{Dtot}, we see
\eq
\frac{\Dtot}{\Deff}\sim e^{\{-(1-\rho)\log(1-\rho)\}L}=e^{\{\rho +O(\rho^2)\}L},
\en
and hence $\Deff$ is considerably smaller compared with the total dimension $\Dtot$.
This conclusion is consistent with the numerical result in \cite{RigolFitzpatrick2011}.
We conclude that our strategy of the proof of Theorem~\ref{t:main} is ineffective for this initial state.
Interestingly, it was found numerically in \cite{RigolMuramatsuOlshanii2006,RigolFitzpatrick2011} that the free fermion chain with similar initial states exhibits thermalization in some sense.


\subsection{$\Deff$ in Golomb-ruler configurations}
\label{s:Golomb}
We next discuss a class of particle configurations for which the effective dimension $\Deff$ is easily evaluated and turns out to be almost as large as the total dimension $\Dtot$.
In these settings, however, the particle density inevitablyapproaches zero as $N$ gets larger.

A sequence of natural numbers $\bsx=(x_1,\ldots,x_N)$ is called a Golomb ruler \cite{GR} if for any $j\ne k$, one has $x_j-x_k=x_\ell-x_m$ only when $j=\ell$ and $k=m$.
The periodic boundary counterpart (in which one replaces the condition $x_j-x_k=x_\ell-x_m$ by $x_j-x_k=x_\ell-x_m \mod L$) is called a modular Golomb ruler.
The optimal (minimum) system size $L$ of a modular Golomb ruler for given $N$ is $L=N(N-1)$, since $x_j-x_k$ takes $N(N-1)$ distinct positive integers.
The optimal configuration, if exists, is called a perfect difference set.
Interestingly, perfect difference sets are proven to exist if $N-1$ is a prime power $p^n$~\cite{Singer}.

We set the configuration of $N$ particles as a modular Golomb ruler $\bsx=(x_1,\ldots,x_N)$ ($x_1<x_2\cdots <x_N$).
By taking $x_1=1$ and choosing the system size $L$ as a prime such that $L\ge2x_N-1$, we see that for any $j\ne k$, one has $x_j-x_k=x_\ell-x_m\mod L$ only when $j=\ell$ and $k=m$ (i.e., a modular Golomb ruler).
A nontrivial and asymptotically optimal example\footnote{
A non-optimal but simple example of a Golomb ruler is obtained by taking $N$ such that $L=2^N-1$ is a (Mersenne) prime, and setting $x_j=2^{j-1}$ for $j=1,\ldots,N$.
In this case, the particle density $\rho\simeq N/2^N$ is exponentially small in $N$.
}
of a Golomb ruler can be found in \cite{ErdosTuran}, where the following sequence
\eq
x_j=1+2N(j-1)+\{(j-1)^2\mod N\},
\lb{ET}
\en
for $j=1,\ldots,N$ with a prime $N>2$ is shown to be a Golomb ruler.
Since $1+2N(N-1)\le x_N\le 1+2N(N-1)+N-1$, the aforementioned construction leads to the chain length as $L\simeq 4N^2$ with the particle density $\rho\simeq(4N)^{-1}$.
Note that the optimal (densest) Golomb ruler has density $\rho=O(N^{-1})=O(L^{-1/2})$, and thus the above construction is asymptotically optimal.

\medskip
We shall fix an arbitrary initial particle configuration $\bsx$ that forms a modular Golomb ruler and evaluate its effective dimension.
We first bound the sign factor in \rlb{Sx} as $(-1)^{QP'Q'}\le1$ to get
\eq
\calS_{\bsx}\le\sumtwo{Q,P',Q'}{(Q\ne\mathrm{id},\,P'\ne Q')}
\prod_{j=1}^N\chi[x_{j}-x_{Q(j)}-x_{P'(j)}+x_{Q'(j)}=0\mod L].
\lb{SxU}
\en
In fact, it can be shown that this is an equality\footnote{%
In a Golomb ruler, $x_{j}-x_{Q(j)}-x_{P'(j)}+x_{Q'(j)}=0\mod L$ holds only if (1)~$j=P'(j)$ and $Q(j)=Q'(j)$, or (2)~$j=Q(j)$ and $P'(j)=Q'(j)$.
Now we decompose a set $\{1,2,\ldots , N\}$ into two subsets, $A$ and $B$, where (1) holds in $A$ and (2) holds in $B$.
Then, $P'$, $Q$, $Q'$ can be expressed in the form of $P'=\mathrm{id}^A\oplus \pi^A$, $Q=\pi^B\oplus \mathrm{id}^B$, and $Q'=\pi^A\oplus \pi^B$, where  $\pi^A$ and $\pi^B$ represent permutations on $A$ and $B$, respectively.
With the above form of permutations, we easily see $(-1)^{P'QQ'}=1$ if $x_{j}-x_{Q(j)}-x_{P'(j)}+x_{Q'(j)}=0\mod L$ holds for any $j$.
}, but the upper bound is enough for our purpose.

Let us fix a permutation $Q\ne\mathrm{id}$, and examine the conditions for $\prod_{j=1}^N\chi[\cdots]=1$, i.e., $x_{j}-x_{Q(j)}-x_{P'(j)}+x_{Q'(j)}=0\mod L$ for all $j=1,\ldots,N$.
If $j$ is such that $Q(j)\ne j$, the condition for $\bsx$ implies $P'(j)=j$ and $Q'(j)=Q(j)$.
We see there is no choice for $P'(j)$ and $Q'(j)$.
If $j=Q(j)$, on the other hand, the only requirement is $P'(j)=Q'(j)$.
There is some freedom for choosing $P'(j)$ and $Q'(j)$.

Let $n_Q$ be the number of $j$ such that $Q(j)=j$.
Since $Q\ne\mathrm{id}$, we see $n_Q=0,1,\ldots,N-2$.
From the above consideration, we see that there are $n_Q!$ choices for $P'$ (and thus $Q'$) for fixed $Q$.
We thus find
\eq
(\text{RHS of \rlb{SxU}})=\sumtwo{Q}{(Q\ne\mathrm{id})}n_Q!=\sum_{n=0}^{N-2}n!\,\calN(n),
\lb{RHSSx}
\en
where $\calN(n)$ is the number of $Q\ne\mathrm{id}$ such that $n_Q=n$.
The value of $\calN(n)$ is computed explicitly as 
\eq
\calN(n)=\binom{N}{n}d_{N-n}, 
\en
where $d_m$ is the $m$-th de Montmort number (also known as the $m$-th derangement number or the subfactorial of $m$).
The de Montmort number counts the number of derangement\footnote{
A derangement is a permutation in which no entry stays at the original position.
}
on $n$ elements.
Fortunately, the de Montmort number $d_m$ is explicitly computed as 
\eq
d_m=\left\lfloor \frac{m!+1}{e}\right\rfloor
\en
with the floor function\footnote{
The floor function $\lfloor x\rfloor$ is the largest integers less than or equal to $x$.
}
$\lfloor \cdot \rfloor$~\cite{Montmort}.
This expression, with \rlb{SxU} and \rlb{RHSSx}, leads to a simple upper bound
\eq
\calS_x\le\sum_{n=0}^{N-2} \frac{N!}{e}\biggl(1+\frac{1}{(N-n)!}\biggr)
=\frac{N!}{e}\biggl(N-1+\sum_{m=2}^N\frac{1}{m!}\biggr)
\le\frac{N!}{e}(N-3+e).
\en
Substituting this into \rlb{DeffS}, we can bound the effective dimension from below as
\eq
\Deff\ge\frac{eL^N}{(N+2e-3)N!}.
\en
Thus the ratio between the total dimension and the effective dimension is bounded as
\eq
\frac{\Dtot}{\Deff}\le \frac{(N+2e-3)\,L!}{e(L-N)!\,L^N}\le \frac{N+2e-3}{e}.
\lb{DDGolomb}
\en
Note that $\Dtot=\binom{L}{N}$ is approximated by $(L/N)^N$ when $N\ll L$, and hence grows super-exponentially in $N$.
(If we take the initial configuration \rlb{ET} then $\Dtot\sim(4N)^N$.)
This means that $\Deff$ satisfying \rlb{DDGolomb} is extremely close to $\Dtot$.

As we have stressed, such a large effective dimension is expected in a non-integrable quantum many-body system, but not in an integrable system.
Here we have a large $\Deff$ in a free fermion model because of the artificial Golomb-ruler configuration.
But recall that this choice is possible only in the extremely low density $\rho=O(N^{-1})$.

\medskip
The above observation about the large effective dimension leads to a statement about thermalization.
Take a sufficiently large and arbitrary prime $N$ and a prime $L$ such that $L\ge2x_N-1$ with $x_N$ given by \rlb{ET}.
We consider the system of $N$ fermions on the chain $\{1,\ldots,L\}$ with the Hamiltonian \rlb{H}.  
We take the phase factor $\theta$ for which the energy eigenvalues \rlb{Ek} are nondegenerate.
(See Lemma~\ref{T:nondeg}.)

For simplicity we restrict our observable only to the particle number in the left half of the chain, i.e.,
\eq
\hN_\mathrm{left}=\sum_{j=1}^{(L+1)/2}\hn_j.
\en
The equilibrium value is of course $\sbkt{\hN_\mathrm{left}}_\infty=N/2$.
Let the initial state be $\ket{\Phi(0)}=\ket{\Phi_{\bsx}}=\hcd_{x_1}\cdots\hcd_{x_N}\vac$, where the configuration $x_1,\ldots,x_N$ is given by \rlb{ET}.
Since
\eq
\frac{\hN_\mathrm{left}}{N}\ket{\Phi(0)}=\ket{\Phi(0)},
\en  
the initial state is highly nonequilibrium with respect to the quantity $\hN_\mathrm{left}/N$.

Then by using the large deviation type estimate
\eq
\biggl\langle\hat{P}\biggl[
\Bigl|\frac{\hN_\mathrm{left}}{N}-\frac{1}{2}\Bigr|\ge\ep
\biggr]\biggr\rangle_\infty\le e^{-(4\ep^2/3)N},
\en
which follows from \rlb{PC}, and the standard argument (as in the proof of Theorem~\ref{t:main}), we can prove the following.
\begin{theorem}
For any $\ep>0$, there exists a constant $T>0$ and a subset (a collection of intervals) $G\subset[0,T]$ with
\eq
\frac{\mu(G)}{T}\ge 1-e^{-(\ep^2/4)N},
\lb{TG2}
\en
where $\mu(G)$ is the total length of the intervals in $G$.
Suppose that one performs a measurement of the number operator $\hN_\mathrm{left}$ in the state $\kPt$ with arbitrary $t\in G$.
Then, with probability larger than $1-e^{-(\ep^2/4)N}$, the measurement result $N_\mathrm{left}$ satisfies
\eq
\Bigl|\frac{N_\mathrm{left}}{N}-\frac{1}{2}\Bigr|\le\ep,
\en
i.e., the state is found in thermal equilibrium.
\end{theorem}
Thus thermalization starting from a deterministic initial state has been established without any unproven assumptions.
Here one can choose the precision $\ep>0$ arbitrarily.
But in order to make the factor $e^{-(\ep^2/4)N}$ negligibly small, one must take $N$ large enough, which means that the density becomes lower.

\section{Possible extension to the finite temperature situation}
\label{s:FiniteT}
Throughout the present paper, we only focused on situations where the initial and the final states correspond to infinite temperature thermal states.
See, in particular, section~\ref{s:Natureoffinalstates}.
We believe that our results can be extended to finite temperature settings with extra technical efforts.
Although we do not elaborate on the extension, we here briefly discuss the setting and essential steps in the proof.

We consider the free fermion Hamiltonian \rlb{H0}, \rlb{H}.
Decompose the Hamiltonian as in \rlb{Hdec}, where we choose $\La_1$ as the half-chain $\{1,\ldots,(L-1)/2\}$.
It follows that $\snorm{\Di\hH}=h_0$ is independent of the system size.
Denote by $\ket{\tilde{\Psi}_j}\in\calH_1$ the normalized eigenstate of $\hH_1$ with eigenvalue $\tilde{E}_j$.
For the energy density $u\in(-2,0)$ and the energy width $\Di u>0$, we define the nonequilibrium microcanonical energy shell by 
\eq
\calH_1^u={\rm span}\bigl\{\ket{\tilde{\Psi}_j}\,\Bigl|\, \bigl|\tfrac{\tilde{E}_j}{N}-u\bigr|\le\Di u\bigr\}\subset\calH_1.
\en
Noting that $\hH_2\ket{\tilde{\Psi}_j}=0$, we observe
\eq
\bra{\tilde{\Psi}_j}(\hH-\tilde{E}_j)^2\ket{\tilde{\Psi}_j}=\bra{\tilde{\Psi}_j}\{(\hH_1-\tilde{E}_j)+\Di\hH\}^2\ket{\tilde{\Psi}_j}
=\bra{\tilde{\Psi}_j}\{(\Di\hH)^2\ket{\tilde{\Psi}_j}\le(h_0)^2,
\en
which implies that $\ket{\tilde{\Psi}_j}$ is (with a minor error when $N$ is large) a superposition of $\ket{\Psi_k}$ such that $|E_k-\tilde{E}_j|\lesssim h_0$.
We thus find that any state $\kPz\in\calH_1^u$ and its time-evolution $\ket{\Phi(t)}=e^{-i\hH t}\kPz$ belongs (again, with minor errors when $N$ is large) to the standard microcanonical energy shell
\eq
\calH_{\rm tot}^u={\rm span}\bigl\{\ket{\Psi_j}\,\Bigl|\, \bigl|\tfrac{E_j}{N}-u\bigr|\le\Di u'\bigr\}\subset\calH_{\rm tot},
\en
with $\Di u'>\Di u$.

In the finite temperature setting, we choose initial state $\kPz$ randomly and uniformly from the nonequilibrium energy shell $\calH^u_1$.
The goal is to show that Theorem~\ref{t:main} (with suitable modifications of constants) is valid for the time-evolved state $\kPz$.

Recalling that $\kPz$ (essentially) belongs to $\calH^u_{\rm tot}$, our strategy for the proof will be to properly replace $\calH_1$ and $\calH_{\rm tot}$ in the original proof with $\calH^u_1$ and $\calH^u_{\rm tot}$, respectively.
Let us see how the proof of the most important estimate of the effective dimension, Theorem~\ref{t:Deff}, is modified.
Interestingly, a small modification is sufficient.
Denoting by $\hP^u_1$ the projection onto $\calH^u_1$, and by $D_1^u$ the dimension of $\calH^u_1$, we find
\eqa
\overline{\Deff^{-1}}
&=\sum_{j=1}^{\Dtot}\overline{\bigl|\bPz\hP^u_1\kPj\bigr|^4}
=\frac{2}{D^u_1(D^u_1+1)}\sum_{j=1}^{\Dtot}\snorm{\hP^u_1\kPj}^4
\nl&\le\frac{2}{D^u_1(D^u_1+1)}\sum_{j=1}^{\Dtot}\snorm{\hP_1\kPj}^2\,\snorm{\hP^u_1\kPj}^2
\nl&\le\frac{2}{D^u_1(D^u_1+1)2^N}\Tr[\hP^u_1]=\frac{2}{(D^u_1+1)2^N},
\lb{Deff2mod}
\ena
which is a faithful extension of the key inequality \rlb{Deff23}.
The analog of Theorem~\ref{t:Deff} is proved if we properly estimate the ratio $D^u_{\rm tot}/D^u_1$.
Another nontrivial (but technical) step for the proof of the desired extension of Theorem~\ref{t:main} is the derivation of the large-deviation upper bound \rlb{PC} for the microcanonical average.

\bigskip
\noindent
{\em Conflict of Interest} || The authors declare no conflicts of interest.

\bigskip
\noindent
{\em Data Availability} || All data and information relevant to this study are presented in the paper.

\bigskip
\noindent
{\em Acknowledgement} || 
It is a pleasure to thank Shelly Goldstein, Takashi Hara, Shu Nakamura, and Marcos Rigol for useful discussions.
We also thank Shin Nakano for his patient guidance in number theory, and  Wataru Kai and Kazuaki Miyatani for letting us know of Lemma~\ref{l:Gauss2} and its proof.
N.S. was supported by JSPS Grants-in-Aid for Early-Career Scientists No. JP19K14615, and H.T. by JSPS Grants-in-Aid for Scientific Research No. 22K03474.

\end{document}